\documentclass[journal,twoside,web]{ieeecolor}
\usepackage{generic}
\usepackage{cite}
\usepackage{amsmath,amssymb,amsfonts,bbm,mathtools}
\usepackage{graphicx}
\usepackage{textcomp}
\usepackage{empheq}
\usepackage{eucal}
\usepackage{mathrsfs}
\usepackage{multirow}
\usepackage{tikz}
\usepackage[colorlinks=false]{hyperref}
\usepackage{orcidlink}
\usetikzlibrary{positioning}

\usepackage{float}
\usepackage{algorithm,algorithmicx,algpseudocode}
\algdef{SE}[DOWHILE]{Do}{doWhile}{\algorithmicdo}[1]{\algorithmicwhile\ #1}%

\makeatletter
\let\NAT@parse\undefined
\makeatother

\newcommand{\bluen}[1]{{#1}}

\allowdisplaybreaks
\usepackage{subcaption}

\usepackage{pgfplots}
\pgfplotsset{compat = 1.3}
\pgfplotscreateplotcyclelist{mycolorlist}{%
	blue,every mark/.append style={fill=blue!80!black},mark=*\\%
	red,every mark/.append style={fill=red!80!black},mark=square*\\%
	brown!60!black,every mark/.append style={fill=brown!80!black},mark=otimes*\\%
	black,mark=star\\%
	blue,every mark/.append style={fill=blue!80!black},mark=diamond*}

\newcommand*\widefbox[1]{\fbox{\hspace{2em}#1\hspace{2em}}}
\newcommand{\mcal}[1]{\mathcal{#1}}

\newcommand{\ldef}{\stackrel{\Delta}{=}}

\newcommand{\mb}[1]{\mathbf{#1}}

\newcommand{\tth}[1]{{#1}^{\text{th}}}

\newcommand{\cbrace}[1]{ \left\{ #1 \right\} }

\newcommand{\bm}[1]{
	\begin{bmatrix}
		#1
	\end{bmatrix}
}
\newcommand{\bt}{{\pmb{\theta}}} % bold theta
\newcommand{\real}[1]{\mathbb{R}^{#1}}

\DeclareMathOperator*{\argmax}{arg\,max}

\DeclareMathOperator{\Tr}{Tr} 

\newcommand{\expect}[2]{\mathbb{E}_{#1} \left[ #2 \right]}
\newcommand{\kl}[2]{D_{\text{KL}} \left( #1 \lVert #2 \right) }

\newcommand{\mj}{\mathcal{J}}

\newtheorem{thm}{Theorem}
\newtheorem{lemma}{Lemma}
\newtheorem{prop}[thm]{Proposition}

\newtheorem{remark}{Remark}
\newcommand{\id}{\text{id}}

\def\BibTeX{{\rm B\kern-.05em{\sc i\kern-.025em b}\kern-.08em
		T\kern-.1667em\lower.7ex\hbox{E}\kern-.125emX}}
\begin{document}	
	\title{Load Restoration in Islanded Microgrids:\\
		Formulation and Solution Strategies}	
	\author{Shourya Bose\orcidlink{0000-0002-2081-9545},~\IEEEmembership{Graduate Student Member, IEEE} and Yu Zhang\orcidlink{0000-0001-7889-2676},~\IEEEmembership{Member, IEEE}
	\thanks{S. Bose and Y. Zhang are with the Department of Electrical and
		Computer Engineering at the University of California, Santa Cruz. Emails:
		\texttt{\{shbose,zhangy\}@ucsc.edu}.		
		This work was supported in part by the Faculty Research Grant of UC Santa
		Cruz, Seed Fund Award from CITRIS and the Banatao Institute at the University of
		California, and the Hellman Fellowship.}}
	\maketitle	
	\begin{abstract}
		Adverse circumstances such as extreme weather events can cause
		significant disruptions to normal operation of electric distribution systems (DS), which includes isolating parts of the DS due to damaged transmission equipment. In this paper, we consider the problem of load restoration in a microgrid (MG) that is islanded from the upstream DS. The MG contains sources of distributed generation such as microturbines and renewable energy sources, as well as energy storage systems (ESS). We formulate the load restoration task as a non-convex optimization problem. This problem embodies the physics of the MG by leveraging a branch flow model, while incorporating salient phenomenon in islanded MGs such as the need for internal frequency regulation, 
		%the unbalanced multiphase nature, 
		and complementarity requirements arising in ESS operations. Since the formulated optimization problem is non-convex, we introduce a convex relaxation which can be solved through model predictive control as a baseline method. However, in order to solve the problem considering its full non-convexity, we leverage a policy-learning method called constrained policy optimization, a tailored version of which is used as our proposed algorithm. The aforementioned approaches, along with an additional deep learning method are compared through extensive simulations.
	\end{abstract}
	

%	\begin{figure}[h]
%		\centering
%		\includegraphics[width=\linewidth]{fig/reward33}
%	\end{figure}
%	\begin{figure}[h]
%		\centering
%		\includegraphics[width=\linewidth]{fig/reward141}
%	\end{figure}
%	\begin{figure}[h]
%		\centering
%		\includegraphics[width=\linewidth]{fig/chargedis33}
%	\end{figure}

	\section{Introduction}
	Extreme weather events such as wildfires, hurricanes, and winter storms pose a
	big threat to the reliable operation of electric distribution systems
	(DS)~\cite{HN:2020}. Those events can disrupt the operation of DS by damaging
	electric transmission equipment such as overhead power lines, thereby curbing
	the delivery of electric power. Traditionally, DS have been designed to be
	reliable during nominal operations and in the face of predictable off-nominal
	operating conditions. Recently, a new paradigm of \emph{resilience} is being
	explored by the power engineering community, which posits that a DS must be
	capable of rapidly recovering to a state of nominal operations post extreme
	weather events~\cite{MP-PM:2017}. As localized distribution systems, microgrids
	(MGs) facilitate system resilience, which are equipped with distributed power
	generation including microturbines (MTs) and renewable energy sources (RES). RES
	may contain energy from photovoltaics, wind turbines, geothermal
		sources, etc. The intermittent nature of power production of RESs necessitates
	energy storage systems (ESS). 
	% 	Therefore, MGs also contain energy storage systems (ESS), which can be used
	%store energy during periods of low power demand and inject energy into the MG
	%during periods of high power demand.
	
	Thus, MGs have sources of power generation as well as energy storage. They are an ideal
	candidate for restoring power demand for loads such as residential homes, 
	industries and critical services such as hospitals, especially when they become disconnected (islanding mode) from the
	upstream DS. Coordination of distributed generation sources within
		islanded MGs occurs through a bi-hierarchical control scheme: the
		lower-level \emph{primary control} allows for communication-free fast response
		to disturbances, while the higher level \emph{secondary} \& \emph{tertiary controls} are used to
		generate setpoints for primary control. The latter act over longer timescales
		while leveraging communication infrastructure available to the MG~\cite{SM-DA:2016}.
	Secondary and tertiary control in MGs is achieved with an MG controller (MGC)~\cite{IEEESTD2030-7:2017}, which is a central computer capable of communicating with and controlling generation and storage elements in the MG. Since load restoration involves coordination of sources and loads in the face of constraints arising due to network physics and finite generation capacities, the MGC's algorithm is best posed as an optimization problem~\cite{AGT-NDH:2011}, which can be solved in real-time by the MGC.

		Load restoration, when posed as an optimization problem, features several unique characteristics which inform strategies to solve it. The first challenge arises from the non-convexity of AC power flow (ACPF) equations, which embody the physics of power transfer. Since the ACPF equations are a part of the constraints in load restoration, any solution thereof cannot be certified as globally optimal~\cite{WAB-AG-KIMM-PAT:2013}. In this paper, we use the \emph{DistFlow} equations instead of ACPF equations, since the former is equivalent to the latter in power networks with radial topologies~\cite{MF-SHL:2013}, which is typical of MGs. The \emph{DistFlow} equations, which lend themselves to intuitive convex relaxations, were first introduced in literature by Baran and Wu~\cite{MEB-FFW:1989}. The second challenge involves discrete decisions which must be made as a part of load restoration process. An important example is the principle of ESS \emph{complementarity}, which states that an ESS may not simultaneously charge and discharge at any given time~\cite{AC-DFG:2014}. This principle has conventionally been encoded using nonconvex~\cite{QZ-KD-ZW-FQ-DZ:2021} or integer~\cite{FS-QW-etal:2020} constraints, or enforced through penalty terms in the objective function~\cite{KG-KB-DC-BT:2020}. The third challenge is based on the observation that in islanded MGs, there is no external source of AC frequency regulation and therefore, it must be done using internal sources such as voltage source inverters (VSI). Thus, the dependence of voltages and AC frequency on real and reactive power generation must be modeled as a (possibly convex) constraint~\cite{NN-SA-AB-JM:2019}. The last challenge involves the uncertainty in forecasts of renewable sources, which renders the calculated solution sub-optimal as the quality of forecasts degrades substantially.
		
		As shown in the sequel, formulating load restoration as an
			optimization problem considering the aforementioned challenges results in a
			non-convex nonlinear program (NLP) defined over multiple time steps. The non-convexity implies that there are no
			guarantees on whether a candidate feasible solution is globally
			optimal. Two approaches may be used to remedy this drawback: use of heuristic solution algorithms, or relaxation of the problem
			into a more tractable form. For the latter approach, it is desirable to generate a
			linear or convex relaxation of the problem, which can then be solved through model predictive control (MPC). MPC is a popular technique for
		optimal control of dynamical systems wherein a control task, posed as an
			optimization problem defined over a long time horizon is decomposed into smaller sequential subproblems and solved. MPC has been applied to several applications such as voltage stability assurance~\cite{LJ-RK-NE:2010},
		demand response in industrial loads~\cite{XZ-GH-JZK-IH:2016},  Volt-VAR
		control~\cite{SCD-etal:2019}, and scheduling PV storage
		systems~\cite{TW-HK-SW:2014}. 
		
		The other approach is to use heuristic solution strategies. In this paper,
			we consider reinforcement learning (RL) for this purpose. RL, and deep RL (in which deep neural networks are used to approximate various functions used in RL) are concerned with determining 
			actions which an agent interacting with an environment should take to maximize it accumulated reward.
		Over the last decade, RL has been considered for various power systems applications
		such as Volt-VAR control~\cite{WW-NY-YG-JS:2020}, EV charge
		scheduling~\cite{HL-ZW-HH:2020}, power management in networked
		MGs~\cite{QZ-KD-ZW-FQ-DZ:2021}, and optimal control of ESS in
		MGs~\cite{JD-ZY-DS-CL-XL-ZW:2019}. A specific RL algorithm which can be useful in the current setting is constrained policy optimization (CPO)~\cite{JA-etal:2017}. CPO aims to find a \emph{policy} which takes as input the current system state and outputs the optimal action with respect to the reward, while satisfying multiple constraints on state and action. The policy is represented as a neural network, which can be \emph{trained}~\cite{lecun2015deep} such that its outputs approach optimality. While CPO has recently been considered for power system applications~\cite{QZ-KD-ZW-FQ-DZ:2021,HL-ZW-LL-HH:2021}, significant challenges remain in tailoring the training procedure of CPO to a given application. Tailoring CPO for load restoration is considered in this paper.
		
		We conclude this section by mentioning other strategies used in load restoration literature. Many frameworks for load restoration consider reconfigurable MGs with inelastic loads, wherein the decision variable includes discrete switching actions and dispatch of generation. Such problems may be solved by posing it as a maximum coverage problem~\cite{YX-etal:2018}, or using spanning tree search~\cite{JL-XYM-CCL-KPS:2014}. On the contrary, we consider a fixed-topology radial MG and elastic loads, while focusing on optimum dispatch of MTs, RESs, and ESSs. Such formulations have been solved in literature using stochastic optimization~\cite{FS-QW-etal:2020}, scenario generation and pruning~\cite{GHASEMI2021106873}, and RL aided by power flow simulators~\cite{XZ-etal:2022}. Many formulations also assume additional infrastructure such as mobile ESSs~\cite{GHASEMI-MESS}. We restrict our attention to cases without specialized infrastructure, and do not consider explicit scenario generation: the CPO agent implicitly generates scenarios as it steers the grid along different trajectories during simulation-based training.
		
%		\blue{We conclude this discussion by noting that, different from our usage of the term, `load restoration' is also used in literature to refer to post-islanding load restoration by reconnecting to the upstream DS~\cite{WL-ZL-etal:2013}. Furthermore, we would like to highlight some other approaches to load restoration}, such as risk-limiting
%		strategies~\cite{ZW-etal:2019}, utilization of wide-area monitoring
%		systems~\cite{WL-ZL-etal:2013}, and expert systems~\cite{MT:2008}.
	
	\subsubsection*{Motivation}
	The principal motivation of this paper is to compare performance of a RL-based MG controller for load restoration over conventional optimization-based techniques such as MPC. There are two advantages which RL has over MPC: firstly, neural-network based controllers (alternatively known as \emph{policy}) trained with RL require significantly lower computational resources for implementation as compared to optimization solvers. Secondly, since RL involves learning from exploratory experience, it is possible for the policy to learn mispredictions in forecasts, thereby increasing robustness of generated solutions.  However, the downside to RL is that it produces black-box models which may generate suboptimal or infeasible solutions. To that end, we seek to tailor CPO policy training such that solutions generated are feasible with respect to DistFlow equations and other operational constraints. This would also alleviate the biggest drawback of interior-point methods (IPM) conventionally used to solve MPC: non-convexity and complementarity constraints may lead to ill-conditioning or exponential solve times of IPM-based solvers.
	
	\subsubsection*{Contribution statement}
	In this paper, we consider load restoration for a fixed-topology islanded MG as an optimization problem, in which we incorporate constraints arising from different elements participating in the MG. As a baseline solution strategy we consider MPC, for which we propose a convex relaxation of the problem. However, in order to solve the exact problem \emph{approximately}, we consider CPO, whose training procedure is adapted specifically for load restoration. Finally, we compare MPC and CPO solution quality through simulations, and both these methods are compared to a brute-force learning approach from a dataset.
	
	We make a few standing assumptions for our formulation of load restoration. First, we assume that the MG has a fixed radial topology, and line parameters such as resistance and reactance are known. Second, we assume that the loads are flexible i.e. they can accept a fraction of the load power demanded, and we also assume that the RES output can be fractionally curtailed. Third, we consider the load demands to be static over the entire duration of the restoration procedure, while forecasts of RES outputs vary in time and are assumed to have errors over the same duration.
	
	The main contributions of our work are summarized as follows:
	\begin{enumerate}
%		\item We formulate a mathematical program for load restoration task in
%		islanded MGs, wherein physical constraints of microturbine (MT), ESS, RES, loads, and sources with droop controlled inverters are practically modeled. An extension of the above model to multiphase unbalanced networks is also presented. Our formulation broadly follows the framework presented in~\cite{FS-QW-etal:2020}, except we also model the frequency dependence of power sharing through droop control in island mode.
			\item Alongside existing relaxations for DistFlow equations, we present new convex relaxations for ESS complementarity and the dependence of inverter voltages on reactive power. These relaxations serve to render MPC sub-problems convex, thereby allowing use of convex optimization solvers for solving the same.
		\item Alternatively, in order to solve the unrelaxed load restoration problem, we consider the use of CPO. Training a policy with CPO involves updating learnable weights of the policy using a quadratically constrained optimization problem multiple times for every episode. In Proposition~\ref{th:OptApprox} and Lemma~\ref{lemma:partialDerivative}, we present a tailored method for efficiently calculating coefficients of said optimization problem.
	\item In Section~\ref{sec:simulation}, we compare the relaxed MPC controller with the CPO policy. The comparisons entail verifying the performance of both under erroneous forecasts of RES output. Furthermore, we compare the feasibility gaps accrued due to the MPC relaxation with those resulting from the CPO policy.

 \end{enumerate}
	
	The remainder of the paper is organized as follows.
	Section~\ref{sec:formulation} presents the problem formulation of load
	restoration. Section~\ref{sec:MPC} shows how the problem can be relaxed and
	solved by MPC. Section~\ref{sec:CPO} details our proposed CPO approach along
	with a procedure on how to train the CPO policy.
	Section~\ref{sec:simulation} compares the solutions obtained by the two
	approaches, with a third approach learning the solutions generated by MPC with a deep neural network.
	% Simulations are done for both single phase and multiphase versions of a benchmark system. 
	Section~\ref{sec:summary}
	concludes this work. All detailed proofs of lemmas and propositions are
	deferred to the Appendix.

	\subsubsection*{Notation}
	The notations $\mathbb{R}$, $\mathbb{C}$, $\mathbb{N}$, and $\mathbb{R}_{+}$
	denote the sets of real numbers, complex numbers, natural numbers, and non-negative
	reals respectively. For a complex number $c\in\mathbb{C}$, $\Re(c)$ and $\Im(c)$ denotes its real and imaginary parts. $\text{conv}(\mcal{A})$
	is the convex hull of set $\mcal{A}$. $\kl{p_1}{p_2}$ denotes the KL-divergence
	between probability distributions $p_1$ and $p_2$. For a real number $a\in\real{}$, $[a]^+$ and $[a]^-$ denote $
	\max\{a,0\}$ and $\max\{-a,0\}$ respectively. Boldface variables represent
	vectors and matrices. $\mb{a}^\top$ and $\mb{a}^H$ represent the transpose, and complex conjugate transpose of vector $\mb{a}$, respectively. $\mb{I}^\id_n$ is the identity matrix of size $n \times n$.
	$\mb{a}\preceq \mb{b}$ denotes the elementwise inequality between two vectors. 
	$\mb{A} \otimes \mb{B}$ denotes the Kronecker product of $\mb{A}$ and $\mb{B}$.
	$\left\lVert \mb{a} \right\rVert_2$ denotes the 2-norm of vector $\mb{a}$.
	$\mathcal{N}(\pmb{\mu},\pmb{\Sigma})$ denotes a multivariate Gaussian
	distribution with mean vector $\pmb{\mu}$ and covariance matrix $\pmb{\Sigma}$. For a random variable $X$, $\expect{}{X}$ denotes its expectation.
	%
	% 	\nopagebreak
	\section{Load Restoration Problem Formulation}\label{sec:formulation}
	Consider an MG that is islanded from the upstream DS due to an extreme weather
	event. Let the MG be represented by a directed graph $\mcal{G} =
	(\mcal{N},\mcal{E})$, where $\mcal{N}$ represents the set of buses and
	$\mcal{E}$ is the set of power lines. We assume that $\mcal{G}$ is a
	\emph{radial network}, i.e., $\mcal{G}$ is a tree. $\mcal{N}$ is the union of
	disjoint sets $\mcal{N}^{\text{L}}$, $\mcal{N}^{\text{MT}}$,
	$\mcal{N}^{\text{RES}}$, and $\mcal{N}^{\text{ESS}}$, where
	$\mcal{N}^{\text{L}}$ represents the load buses, $\mcal{N}^{\text{MT}}$
	represents the buses connected to an MT, $\mcal{N}^{\text{RES}}$ represents the
	buses with RES, and $\mcal{N}^{\text{ESS}}$ represents the buses connected to an
	ESS. We consider the load restoration problem over a time horizon of $\mcal{T}
	\ldef \{1,2,\cdots,T\}$, and time steps are indexed by $t$. We let
	$s_{i,t}\in\mathbb{C}$ and $v_{i,t}\in\mathbb{R}_+$ denote the complex power injection and squared magnitude of voltage phasor at bus $i\in\mcal{N}$
	respectively at time $t$. For every line $(i,j)\in\mcal{E}$, we let $S_{ij,t}\in\mathbb{C}$, and $l_{ij,t}\in\mathbb{R}_+$ denote the \emph{sending-end} complex power flow and squared magnitude of the current phasor, respectively. For buses $i$
	and $j$, notation $i\rightarrow j$ indicates the presence of a power line in
	between, i.e. $(i,j)\in \mcal{E}$. Note that any equation involving the index $t$, unless stated otherwise, is assumed to hold for all $t\in\mcal{T}$.
	% 	Note that the power terms $P_{i,t},Q_{i,t},Q_{ij,t},Q_{ij,t}$ assume values
	%in $\mathbb{R}$, while the squared voltage magnitude $v_{i,t}$ and current
	%magnitude $l_{ij,t}$ take nonnegative real values. Furthermore, a positive
	%value of any power quantity represents injection into the MG, while a negative
	%value represents withdrawal from the MG.
	
	\subsubsection*{Objective function}
	The objective is to maximize load restoration while minimizing MT fuel
	consumption. To that end, an appropriate objection function can be represented as
	\begin{equation}
		\label{eq:objective}
		J_\mcal{T} \ldef\sum_{t\in\mcal{T}} \left(  \sum_{i\in\mcal{N}^\text{L}} C^\text{L}_{i,t}(\Re(s_{i,t})) + \sum_{i\in\mcal{N}^\text{MT}} C^{\text{MT}}_{i,t} (\Re(s_{i,t}))   \right),
	\end{equation}
	where the (possibly time-varying) functions $C^L$ and $C^{\text{MT}}$ are concave functions considering the sign convention that generated power is positive while consumed power is negative. The objective function is meant to incentivize the amount of load restored, while disincentivizing power produced from MT in favor of utilizing ESSs or RESs. In practice, $C^L$ is often linear with coefficients representing priority order of load restoration, while $C^{\text{MT}}$ can be linear or concave quadratic.
	
	\subsubsection*{Power flow constraints}
	Let $z_{ij}\in\mathbb{C}$ represent the impedance of line
	$i \rightarrow j$. We use the \emph{DistFlow} equations for quantifying the
	power flows that are given as follows:
	\begin{subequations}
		\label{eq:PF}
		\begin{align}
			\label{eq:PFcons1}
			s_{j,t} &= \sum_{j\rightarrow k} S_{jk,t} - \sum_{i\rightarrow j} \left( S_{ij,t} - z_{ij}l_{ij,t} \right), \forall j \in\mcal{N}\\
			\label{eq:PFcons2}
			v_{j,t} &= v_{i,t} - 2\Re(\bar{z}_{ij}S_{ij,t}) + |z_{ij}|^2l_{ij,t},\, \forall i \rightarrow j\\
			\label{eq:PFcons3}
			v_{i,t}l_{ij,t} &= |S_{ij,t}|^2, \; \forall i \rightarrow j.
		\end{align}
	\end{subequations}
	It is known that in radial networks, \bluen{voltage angles can be recovered from any solution of equations~\eqref{eq:PFcons1} to~\eqref{eq:PFcons3}~\cite[Theorem 2]{MF-SHL:2013}. The voltage and current constraints at each bus and line respectively, needed for nominal operation of the MG, are codified as follows:}
	\begin{gather}
		\label{eq:voltagecons}
		\underline{v} \leq v_{i,t} \leq \bar{v}, \; \forall i\in\mcal{N},\\
		\label{eq:CurrentConstraint}
		l_{ij,t} \leq \bar{l}_{ij}, \; \forall i \rightarrow j.
	\end{gather}
	\subsubsection*{MT Constraints}
	The power generation of each MT is subject to constraints on per-time step generation, as well as those on rates of ramping up/down. They are given as
	\begin{subequations}
		\label{eq:MT_lim_and_ramp}
		\begin{gather}
			\underline{P}^\text{MT}_i \leq \Re(s_{i,t}) \leq \bar{P}^\text{MT}_i, \; \forall i
			\in\mcal{N}^\text{MT},\\
			\label{eq:MT_lin_and_ramp_2}
			\underline{P}^\text{MT}_\text{rd} \leq \Re(s_{i,t}) - \Re(s_{i,t-1}) \leq
			\bar{P}^\text{MT}_\text{ru}, \forall i \in \mcal{N}^\text{MT},\\
			\Re(s_{i,1})  \leq \bar{P}^\text{MT}_\text{ru}, \; \forall i \in \mcal{N}^\text{MT},
		\end{gather}
	\end{subequations}
	wherein~\eqref{eq:MT_lin_and_ramp_2} holds for all $t\in\mcal{T}\setminus \{1\}$.
	Each MT is assumed to have a fixed amount of fuel at the beginning of
	$\mcal{T}$. This constrains the total amount of real power produced over
	$\mcal{T}$ as $\left( \sum_{t\in\mcal{T}} \Re(s_{i,t}) \right) \tau_i \leq E_i, \; \forall i \in
	\mcal{N}^\text{MT}$, wherein $E_i$ is the total fuel initially available at MT $i$, and $\tau_i$ is
	the power-to-fuel conversion factor. The
	total fuel constraint can be reformulated as a recursive relation by denoting with $\zeta_{i,t}$ the
	amount of fuel remaining in MT $i$ at time $t$ and observing that
	\begin{subequations}
		\label{eq:MTfuelalt}
		\begin{gather}
			\label{eq:MTfuelalt1}
			\zeta_{i,t} = \zeta_{i,t-1} - \tau_i \Re(s_{i,t}), \; \forall
			i\in\mcal{N}^\text{MT}\\
			\label{eq:MTfuelalt2}
			\zeta_{i,T}\geq 0, \; \forall i \in \mcal{N}^\text{MT}
		\end{gather}
		with the initial condition $\zeta_{i,0} = E_i$.
	\end{subequations}
	\subsubsection*{ESS Constraints}
	Each ESS is an energy reservoir which may discharge power into the MG when
	required, and otherwise charge in order to replenish its energy reserves. We denote by
	$\mcal{S}_{i,t}$ the state of charge (SoC) of ESS $i$ at time step $t$, which takes values in
	$[\underline{\mcal{S}}_i,\bar{\mcal{S}}_i]$. Letting $P^\text{ch}_{i,t}$ and
	$P^\text{dis}_{i,t}$ denote the charge and discharge powers of the same, the power input/output constraints of the ESS are:
	\begin{subequations}
		\label{eq:OrigESSModelandInj}
		\begin{gather}
			\label{eq:charge}
			0 \leq P^\text{dis}_{i,t} \leq \bar{P}_i^\text{dis}, \; 0 \leq P^\text{ch}_{i,t} \leq \bar{P}_i^\text{ch}, \; \forall i \in
			\mcal{N}^\text{ESS},\\
			\label{eq:complementarity}
			P^\text{ch}_{i,t}P^\text{dis}_{i,t} = 0, \; \forall i \in
			\mcal{N}^\text{ESS},\\
			\label{eq:ESSNetRealInj}
			\Re(s_{i,t}) =P^{\text{dis}}_{i,t} -  P^{\text{ch}}_{i,t}, \; \forall i
			\in\mcal{N}^\text{ESS}.
		\end{gather}
	\end{subequations}
	Note that \eqref{eq:complementarity} denotes the complementarity constraint, which may alternatively be denoted as an integer constraint using a charge/discharge indicator variable. The evolution of $\mcal{S}_{i,t}$ is given as:
	\begin{subequations}
		\label{eq:SoCcons}
		\begin{gather}
			\label{eq:SoCcons1}
			\mcal{S}_{i,t} = \mcal{S}_{i,t-1} + \eta^\text{ch}_iP^\text{ch}_{i,t}\Delta_t - \frac{1}{\eta^\text{dis}_i} P^\text{dis}_{i,t} \Delta_t, \; \forall i \in \mcal{N}^\text{ESS}, \\
			\mcal{S}_{i,0} = \mcal{S}^\text{init}_{i}, \; \mcal{S}_{i,t} \in
			[\underline{\mcal{S}}_i,\bar{\mcal{S}}_i], \; \forall i \in \mcal{N}^\text{ESS},
		\end{gather}
	\end{subequations}
	wherein~\eqref{eq:SoCcons1} holds for all $t\in\mcal{T}$, $\eta^\text{ch}_i\in(0,1]$ and $\eta^\text{dis}_i \in (0,1]$ denote the
	charge and discharge efficiency for ESS $i$, $\Delta_t$ denotes the time
	duration corresponding to each time step, and $\mcal{S}^\text{init}_i$ denotes
	the initial SoC of ESS $i$. Considering other factors such as battery temperature, the efficiencies $\eta^{\text{ch}}_i$ and $\eta^{\text{dis}}_i$ can be modeled as time-varying~\cite{LWC-MFMS-ABI-ZFH:2004}. 
	\subsubsection*{RES Constraints}
	Since the power output of RES is stochastic in nature and cannot be predicted with
	certainty ahead-of-time, a forecast
	$\hat{P}^\text{RES}_{i,t}$ is used as a stand-in for the actual output. We assume each RES has the capability to curtail its power output, and denote by $\kappa_{i,t}\in[0,1]$ the curtailment
	ratio of the real power. The RES real power constraint is given as
	\begin{equation}
		\Re(s_{i,t}) = (1-\kappa_{i,t})\hat{P}^\text{RES}_{i,t}, \; \forall i \in
		\mcal{N}^\text{RES}.
	\end{equation}
	\subsubsection*{Reactive Power and Droop Bus Constraints}
	The MT, RES, and ESS buses may be interfaced to the MG through inverters which convert DC to AC power. Such inverters are capable of supplying and absorbing reactive power to and from the MG, constrained as
	\begin{equation}
		\label{eq:ReacPowerCons}
		|\Im(s_{i,t})| \leq \bar{Q}_{i}, \; \forall i \in \mcal{N}^\text{MT} \cup
		\mcal{N}^\text{ESS} \cup \mcal{N}^\text{RES},
	\end{equation}
	where $\bar{Q}_i$ is the nameplate capacity of the inverter at bus $i$. Other representations of inverter capacity that limit the total apparent power of the inverter (by way of an upper bound on $|s_{i,t}|$) may also be used in lieu of~\eqref{eq:ReacPowerCons}.
	
	Many inverters in MGs may operate on the principle of \emph{droop control}, also referred to as \emph{voltage source inverter control}, wherein the inverter acts as an AC voltage source, with voltage and frequency depending on the real and reactive power output. Letting $\mcal{N}^\text{droop} \subset \left\{\mcal{N}\setminus\mcal{N}^\text{L} \right\}$ denote the set of generation buses operating on droop control (`droop buses'), and $\omega_t$ denote the system frequency at time $t$, the droop bus constraints are given as~\cite{JAPL-CLM-AGM:2006}:
		\begin{subequations}
			\label{eq:DroopCons}
			\begin{gather}
				\omega_{t} = \omega_i^* - k_P(\Re(s_{i,t}) - P^*_i), \; \forall i \in \mcal{N}^\text{droop}\\
				\label{eq:droop_reactive}
				\sqrt{v_{i,t}} = \sqrt{v^*_i} - k_Q(\Im(s_{i,t}) - Q^*_i), \; \forall i \in \mcal{N}^\text{droop},\\
				\label{eq:freq_regulation}
				\underline{\omega} \leq \omega_t \leq \bar{\omega},
			\end{gather}
		\end{subequations}
		wherein $\omega^*$, $v_i^*$, $P_i^*$, and $Q_i^*$ are operational setpoints of the inverter, while $k_P$ and $k_Q$ are the droop constants. Since the MG receives no frequency signals from the upstream DS during islanding, droop buses are essential for internal frequency regulation of the system, which is achieved through constraint~\eqref{eq:freq_regulation}.
	
%	\orange{Note that for the current work, we assume that only droop buses are capable of setting the MG frequency. However, frequency control may be achieved by MT, ESS, RES buses~\cite{NL-CZ-LC:2016}, as well as load buses~\cite{CZ-UT-NL-SL:2014}. Further, apart from droop control, distributed optimization frameworks~\cite{MA-DVD-KHJ-HS:2013}, for example consensus algorithms~\cite{LX-YM-YCT-GL-HS-CP-MF:2019} have also been considered for frequency control in islanded MGs. However, in order to maintain simplicity and derive a relaxation of the droop equations in Lemma~\ref{lemma:DroopBus}, we restrict our attention to frequency setting through droop buses.}
	%
	\subsubsection*{Load Constraints}
	We let time-varying forecasts of active and reactive power demands of the loads be denoted by $\hat{P}^\text{L}_{i,t}$ and
	$\hat{Q}^\text{L}_{i,t}$  respectively. We let $\rho_{i,t}\in[0,1]$ denote the
	pickup ratio (i.e. ratio of load served with respect to demand) of load $i\in\mcal{N}^\text{L}$ at time $t$. Assuming a constant power factor, we have the constraints
	\begin{align}
		\Re(s_{i,t}) = -\rho_{i,t} \hat{P}^\text{L}_{i,t},\, \; \Im(s_{i,t}) = -\rho_{i,t}
		\hat{Q}^\text{L}_{i,t},\, \forall i \in \mcal{N}^\text{L}.
	\end{align}
	A monotonically increasing pickup ratio is preferable over frequent dropping of
	loads already picked up. This is ensured by the \emph{almost-monotonic} load restoration constraint
	\begin{align}
		\label{eq:load_monotonicity}
		\rho_{i,{t}} -\rho_{i,t-1} \geq -\epsilon,\; \forall i\in\mcal{N}^\text{L},
		\forall t \in \mcal{T} \backslash \{1\},
	\end{align}
	where parameter $\epsilon\geq 0$ allows for a small leeway in monotonicity
	of load pickup, and is chosen by the system operator.
	\subsubsection*{Optimization Problem}
	We collect all decision variables at time $t$ as
	\begin{align*}
		\mcal{X}_t \ldef& \big\{ \cbrace{s_{i,t},v_{i,t}}_{i\in\mcal{N}},\cbrace{S_{ij,t},l_{ij,t}}_{i\rightarrow j},\cbrace{\zeta_{i,t}}_{i\in\mcal{N}^\text{MT}},\\
		& \cbrace{\mcal{S}_{i,t},P^\text{ch}_{i,t},P^\text{dis}_{i,t}}_{i\in\mcal{N}^\text{ESS}}, \cbrace{\kappa_{i,t}}_{i\in\mcal{N}^\text{RES}}, \cbrace{\rho_{i,t}}_{i\in\mcal{N}^\text{L}},\omega_{t} \big\}.
	\end{align*}
	To this end, the optimization problem to be solved by the MGC is given as
	\begin{subequations}
		\label{eq:optprob}
		\begin{empheq}[box=\widefbox]{align}
			\max\limits_{\cbrace{\mcal{X}_t}_{t\in\mcal{T}}} \quad &\eqref{eq:objective}\\
			\text{s.t.}\quad &\eqref{eq:PF}- \eqref{eq:load_monotonicity}.
		\end{empheq}     
	\end{subequations}
	Obtaining an optimal solution to~\eqref{eq:optprob} allows the MGC to implement said solution for load restoration. Note that the constraints~\eqref{eq:PFcons3},~\eqref{eq:complementarity}, and~\eqref{eq:droop_reactive} are non-convex, thereby making~\eqref{eq:optprob} overall non-convex and devoid of global optimality guarantees.

	\section{Convex Relaxation and Solution using Model Predictive Control}\label{sec:MPC}
	In this section, we consider convex relaxations to problem~\eqref{eq:optprob}, which
	make it amenable to MPC. The MPC approach involves
	solving~\eqref{eq:optprob} over subhorizons of $\mcal{T}$, followed by using the
	subhorizon-optimal solutions to construct a near-optimal solution over
	$\mcal{T}$~\cite{JBR:2000}. The non-convexity arising from DistFlow equations~\eqref{eq:PFcons3} can be addressed using a well-studied second-order cone relaxation $v_{i,t}l_{ij,t} \geq |S_{ij,t}|^2$, equivalently written as a second-order cone, 
	\begin{align*}
		\left\lVert \bm{2\Re(S_{ij,t}) & 2\Im(S_{ij,t}) & l_{ij,t} - v_{i,t} }^\top \right\rVert_2 \leq l_{ij,t} + v_{i,t}.
	\end{align*}
	We use the above relaxation for the MPC formulation. 
	%The multiphase representation given by~\eqref{eq:ACDF3}, can be rendered convex simply by dropping the rank-1 constraint.
	Next, we consider the droop bus equation~\eqref{eq:droop_reactive} which is non-convex due to the presence of the term $\sqrt{v_{i,t}}$. It can be relaxed to a second-order cone as follows.
	\begin{lemma}[Convex relaxation of~\eqref{eq:droop_reactive}]\label{lemma:DroopBus}
		The constraint $\sqrt{v_{i,t}} = \sqrt{v_i^*} - k_Q(\Im(s_{i,t})-Q_i^*)$ can be relaxed to a second-order cone given as
		\begin{align*}
			\left\lVert \bm{ \left(\sqrt{v_i^*} -k_Q(\Im(s_{i,t})-Q_i^*)\right) & v_{i,t} & \frac{1}{2}}^\top \right\rVert_2 \leq v_{i,t}+\frac{1}{2}.
		\end{align*}
	\end{lemma}
	\begin{proof}
		See Appendix ~\ref{proofLemma2}.
	\end{proof}
	Lastly, we consider the non-convex ESS complementarity constraints (CC). 
	% In~\eqref{eq:optprob}, the feasible sets of $P^\text{ch}_{i,t}$ and $P^\text{dis}_{i,t}$ for all $i$ and $t$ as described by constraints in~\eqref{eq:OrigESSModelandInj} are non-convex due to the presence of CC~\eqref{eq:complementarity}. 
	We address the non-convexity by relaxing the joint feasible region of $P^\text{ch}_{i,t}$ and $P^\text{dis}_{i,t}$ to its convex hull, whose closed form is given as follows.
	%any given set.
	\begin{lemma}[Convex hull of nonconvex ESS CC feasible
		set]\label{lemma:ConvexHull}
		Define the set $\mcal{P}_i^t \ldef \cbrace{ (P^\text{ch}_{i,t},P^\text{dis}_{i,t}) \, | \,
			\eqref{eq:charge}-\eqref{eq:complementarity}}$.
		Then, the convex hull of $\mcal{P}_i^t$ is given as
		\begin{align*}
			\mcal{P}_i^{t,\text{conv}} =& \cbrace{ (P^\text{ch}_{i,t},P^\text{dis}_{i,t}) \,\bigg| \,
				P^\text{ch}_{i,t}\geq 0, P^\text{dis}_{i,t} \geq 0,  \frac{P^\text{ch}_{i,t}}{\bar{P}^\text{ch}_i}
				+ \frac{P^\text{dis}_{i,t}}{\bar{P}^\text{dis}_i} \leq 1}.
		\end{align*}
	\end{lemma}
	\begin{proof}
		See Appendix~\ref{proofLemma1}.
	\end{proof}
	The relaxation of CC presented in Lemma~\ref{lemma:ConvexHull} replaces the constraints~\eqref{eq:OrigESSModelandInj}. A concern is that the relaxed solution may be physically unimplementable due to non-complementarity. However, based on results presented in~\cite{NN-MA:2021}, it is possible to recover a charge/discharge schedule which results in the same power injections~\eqref{eq:ESSNetRealInj}, while satisfying CC. For an ESS $i\in\mcal{N}^\text{ESS}$, consider a charge/discharge schedule satisfying constraints in Lemma~\ref{lemma:ConvexHull}, given as $\mb{P}_i^{\text{ch}} \ldef \{P_{i,t}^\text{ch} \}_{t\in\mcal{T}}$ and $\mb{P}_i^{\text{dis}} \ldef \{P_{i,t}^\text{dis} \}_{t\in\mcal{T}}$. Consider an alternate schedule given as $\hat{\mb{P}}^\text{ch}_i\ldef \{\hat{P}^\text{ch}_{i,t} \}_{t\in\mcal{T}}$ and $\hat{\mb{P}}^\text{dis}_i\ldef \{\hat{P}^\text{dis}_{i,t} \}_{t\in\mcal{T}}$, such that 
		$\hat{P}^\text{ch}_{i,t} \ldef [P^\text{ch}_{i,t}-P^\text{dis}_{i,t}]^+,\quad \hat{P}^\text{dis}_{i,t} \ldef [P^\text{dis}_{i,t}-P^\text{ch}_{i,t}]^+$.
	The alternate schedule results in the same net injection, since 
	\begin{align*}
		\hat{P}^\text{dis}_{i,t} - \hat{P}^\text{ch}_{i,t} = [P^\text{dis}_{i,t}-P^\text{ch}_{i,t}]^+ - [P^\text{ch}_{i,t}-P^\text{dis}_{i,t}]^+ = P^{\text{dis}}_{i,t} - P^\text{ch}_{i,t}
	\end{align*}
	Furthermore, if $(P^\text{ch}_{i,t},P^\text{dis}_{i,t})\in\mcal{P}^{t,\text{conv}}_{i,t}$, then it can be verified that $\hat{P}^\text{ch}_{i,t}$ and $\hat{P}^\text{ch}_{i,t}$ satisfy~\eqref{eq:charge}-\eqref{eq:complementarity}.
	However, the alternate charging/discharging schedule leads to the original SoC being an underestimate, which we show through induction. Let $\hat{\mcal{S}}_{i,t}$ be the SoC at time step $t$ under the schedule $\hat{\mb{P}}^\text{ch}_i$ and $\hat{\mb{P}}^\text{dis}_i$, while $\mcal{S}_{i,t}$ is the SoC under $\mb{P}^\text{ch}_i$ and $\mb{P}^\text{dis}_i$. Further, assume that $ \hat{\mcal{S}}_{i,t}\geq \mcal{S}_{i,t}$ holds for some time step $t$. Since it holds that $P^\text{dis}_{i,t}\geq [P^\text{dis}_{i,t}-P^\text{ch}_{i,t}]^+=\hat{P}^\text{dis}_{i,t}$ and $\eta^\text{ch}_i\eta^\text{dis}_i\leq1$,
	the difference between SoC on subsequent time steps can be expressed as
	\begin{align*}
		&{\mcal{S}}_{i,t+1}-{\mcal{S}}_{i,t} = \Delta_t\left[\eta^\text{ch}_{i}\hspace{-0.2em}\left(  {P}^\text{ch}_{i,t} - P^\text{dis}_{i,t}\right) - \left( \eta_i^\text{ch}-\frac{1}{\eta^\text{dis}_{i}}\right)\hspace{-0.3em} {P}^\text{dis}_{i,t}\right]\\
		\leq&  \Delta_t\left[\eta^\text{ch}_{i}\hspace{-0.2em}\left(  \hat{P}^\text{ch}_{i,t} - \hat{P}^\text{dis}_{i,t}\right) - \left( \eta_i^\text{ch}-\frac{1}{\eta^\text{dis}_{i}}\right)\hspace{-0.3em}\hat{P}^\text{dis}_{i,t}\right] = \hat{\mcal{S}}_{i,t+1}-\hat{\mcal{S}}_{i,t},
	\end{align*}
	and therefore, $\hat{\mcal{S}}_{i,t+1}-{\mcal{S}}_{i,t+1}\geq \hat{\mcal{S}}_{i,t}-{\mcal{S}}_{i,t}$. Combined with the induction assumption, and starting off from the same initial SoC (i.e. $\hat{\mcal{S}}_{i,0}={\mcal{S}}_{i,0}$), it follows that $\hat{\mcal{S}}_{i,t}\geq {\mcal{S}}_{i,t}$ for all $t\in\mcal{T}$. Therefore, we showed that the convex relaxation presented in Lemma~\ref{lemma:ConvexHull} can be converted \emph{post-hoc} into a schedule which respects ESS CC, at the cost of underestimating the ESS SoC.

		\subsubsection*{MPC Implementation}
		\begin{figure}[t!]
			\centering
			\includegraphics[width=0.75\linewidth]{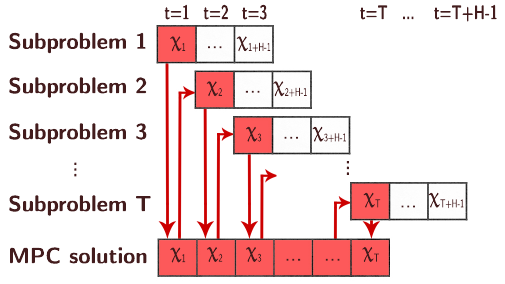}
			\caption{Schematic of load restoration solution with MPC, with look-ahead window $H=3$.}
			\label{fig:mpcschematic}
		\end{figure}
		In this subsection, we briefly sketch out the implementation details of MPC. 
%		For a full treatise on principles of MPC, the reader is referred to~\cite{JBR:2000}. Previously, we considered convex relaxations and/or approximations of~\eqref{eq:optprob}. 
		Even the convexified problem may be difficult to solve over long time horizons, i.e. when $T$ is large, since the number of decision variables scales as $O(T)$. MPC posits that~\eqref{eq:optprob} may be approximately solved by dividing it into $T$ sub-problems, with the $t^\text{th}$ subproblem defined over the variables $\{ \mcal{X}_{t},\cdots,\mcal{X}_{t+H-1} \}$ and considering the relaxed constraints restricted to those involving the variables $\{\mcal{X}_{t},\cdots,\mcal{X}_{t+H-1}\}$. Since $\mcal{X}_{t-1}$ is not a decision variable in the $t^\text{th}$ subproblem, its value is fixed as the optimal value of $\mcal{X}_{t-1}$ derived from the $(t-1)^\text{th}$ subproblem. Thus, MPC allows for a solution of~\eqref{eq:optprob} by solving $T$ subproblems over time horizons of size $H$, instead of one problem over a time horizon of size $T$. The scheme is represented visually in Figure~\ref{fig:mpcschematic}. Note that choosing $H=T$ and considering only the first subproblem reduces to solving the relaxed instance of~\eqref{eq:optprob} in a one-shot fashion.
	
	\section{Solution using Constrained Policy Optimization}\label{sec:CPO}
	
		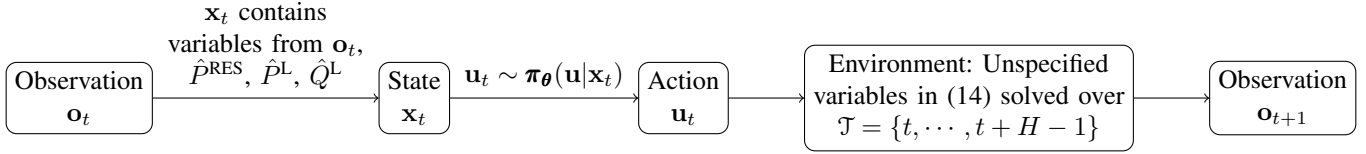
\begin{figure*}[t!]
			\centering
			\begin{tikzpicture}[->,node distance=2cm,auto]
				% Define nodes with rectangles
				\node[draw, rounded corners, align=center] (A) {Observation\\$\mb{o}_{t}$};
				\node[draw, rounded corners, align=center, right=3cm of A] (B) {State\\$\mb{x}_{t}$	};
				\node[draw, rounded corners, align=center, right=2.5cm of B] (C) {Action\\$\mb{u}_t$};
				\node[draw, rounded corners, align=center, right=1cm of C] (D) {Environment: Unspecified\\ variables in~\eqref{eq:optprob} solved over\\$\mathcal{T}=\{t,\cdots,t+H-1\}$};
				\node[draw, rounded corners, align=center, right=1cm of D] (E) {Observation\\$\mb{o}_{t+1}$};
				
				% Draw edges with text
				\path (A) edge node[align=center] {$\mb{x}_{t}$ contains\\variables from $\mb{o}_{t}$,\\ $\hat{P}^\text{RES}$, $\hat{P}^\text{L}$, $\hat{Q}^\text{L}$} (B);
				\path (B) edge node {$\mb{u}_t\sim\pmb{\pi}_{\pmb{\theta}}(\mb{u}|\mb{x}_t)$} (C);
				\path (C) edge node {} (D);
				\path (D) edge node {} (E);
			\end{tikzpicture}
			\caption{Schematic of RL framework to solve load restoration.}
			\label{fig:RLschematic}
	\end{figure*}
	
	We motivate the use of CPO to solve~\eqref{eq:optprob} by considering whether the decision variables and available information for the $t^\text{th}$ MPC subproblem (which starts at time $t$), denoted by $\mb{X}_t \ldef \{\mcal{X}_{t'}\}_{t'=t}^{t+H-1}\cup \mcal{X}_{t-1}$ can be split into three groups, \emph{viz.} state, action, and observation variables, such that they have the following properties:
		\begin{itemize}
			\item The \textbf{state} at time $t$, denoted by $\mb{x}_t$, represents the smallest set of variables in $\mb{X}_t$ plus exogenous data such as forecasts $\hat{P}^\text{RES}$, $\hat{P}^\text{L}$, and $\hat{Q}^\text{L}$, which can adequately describe the system at time $t$.
			\item The \textbf{action} at time $t$, denoted by $\mb{u}_t$, represents the set of variables in $\mb{X}_t$ which the MGC can control.
			\item The \textbf{observation} at time $t$, denoted by $\mb{o}_t$, denotes all variables in $\mb{X}_t \setminus \{ \mb{u}_t \}$. Intuitively, the observation variables should be fully specified once $\mb{x}_t$ and $\mb{u}_t$ are known.
		\end{itemize}
		If such a split is possible, it allows for the introduction of a \emph{policy}, which is a randomized map from a given state to a distribution over possible actions. Should we also define an objective function, which in the RL framework is called the \emph{reward}, it becomes a well-posed problem to seek an optimal policy which maps state to actions in a way which maximizes the total accumulation of reward. Similar to MPC, each action is restricted to the subhorizon $\{t,\cdots,t+H-1\}$, and load restoration is done in a rolling-horizon fashion.

	Armed with the motivation, we pose problem \eqref{eq:optprob} as a constrained Markov
	decision process (CMDP) under the RL policy-learning framework of CPO. A CMDP is
	defined as the 6-tuple $\cbrace{\mathbb{X},\mathbb{U},p,R,\mb{{C}},\gamma}$,
	where $\mathbb{X}$ is the state space, $\mathbb{U}$ is the action space,
	$p:\mathbb{X} \times \mathbb{U} \times \mathbb{X} \mapsto [0,1]$ is the state
	transition probability, $R:\mathbb{X} \times \mathbb{U} \mapsto \mathbb{R}$ is
	the reward function, $\mb{C}:\mathbb{X}\times \mathbb{U} \mapsto
	\mathbb{R}^M$ is the constraint function, and $\gamma\in (0,1]$ is the
	\emph{discount factor} used to de-emphasize the contribution of uncertain future
	quantities to the reward. For many systems including the one under consideration, $p$, or the state transition probabilities are deterministic, based on known rules. The constraint function $\mb{C}$ is used to encode $M\geq 0$ constraints by representing them as $\mb{C}(\mb{x}_t,\mb{u}_t)\preceq \mb{0}$ on every time step $t$. 
%	Note that CPO specifically, and the RL framework in general allows for a \emph{look-ahead window}, which is a smaller subhorizon over which a policy is optimized. Since the basic idea underpinning look-ahead windows is the same as MPC, we will continue to use the variable $H$ to denote the length of the look-ahead window.

	We now describe load restoration in the CPO framework by defining salient variables and functions. Then, we describe a tailored training process to find an optimal policy. Finally, we discuss the issue of evaluating various mathematical expressions which arise during the training process. We define the \emph{state} $\mb{x}_t$ and
	the \emph{action} $\mb{u}_t$ at time $t$ as:
	\begin{align}
		\notag
		\mb{x}_t \ldef& \bigg\{ \cbrace{\mcal{S}_{i,t-1}}_{i\in\mcal{N}^\text{ESS}},
		\cbrace{\zeta_{i,t-1}}_{i\in\mcal{N}^\text{MT}}, \cbrace{\rho_{i,t-1}}_{i\in\mcal{N}^\text{L}},\\
		\label{eq:act_var}
		&\bigg[ \cbrace{ \hat{P}^\text{L}_{i,t'},
			\hat{Q}^\text{L}_{i,t'}}_{i\in\mcal{N}^\text{L}}, \cbrace{
			\hat{P}^\text{RES}_{i,t'}}_{i\in\mcal{N}^\text{RES}}  \bigg]_{t'=t}^{t+H-1}
		\bigg\} \\
		\notag
		\mathbf{u}_t \ldef& \bigg\{ \bigg[ \cbrace{s_{i,t'}}_{i \in
			\mcal{N}^\text{RES}}, \cbrace{s_{i,t'}}_{i\in \mcal{N}^\text{MT}}, \cbrace{\rho_{i,t'}}_{i\in\mcal{N}^\text{L}}\\
		\label{eq:state_var}
		&\cbrace{P^\text{ch}_{i,t'},P^\text{dis}_{i,t'},\Im(s_{i,t'})}_{i\in
			\mcal{N}^\text{ESS}} \bigg]_{t'=t}^{t+H-1} \bigg\}.
	\end{align} 
	The observation variables are simply defined as $\mb{o}_t \ldef \mb{X}_t \setminus \{ \mb{u}_t \}$. 
	A \emph{policy} $\pi_\bt: \mathbb{X} \times \mathbb{U} \mapsto [0,1]$
	parameterized by $\bt\in\mathbb{R}^h$ (where $h$ is the number of elements in the parameter vector) and denoted as
	$\pi_\bt(\mb{u}|\mb{x})$ gives the probability of taking action $\mathbf{u}$
	given the current state $\mb{x}$. Usually, the policy maps from observation to action variables; however, we choose a \emph{fully observable} setup wherein $\mb{x}_t$ contains variables from $\mb{o}_t$ as well as forecasts, and the policy mapping from state to action variables follows. A schematic of the evolution of state, action, and observation variables is shown in Figure~\ref{fig:RLschematic}.
%	There are many ways to model a policy mapping described using a parameter vector~\cite[Section 13.1]{RSS-AGB:2018}. 
	We model the policy as a multivariate Gaussian
	distribution whose mean vector and covariance matrix are generated by a
	feedforward neural network (FNN). This allows the policy to \emph{explore} various possible trajectories during the process of training. To this end, let $d$ denote the dimension of
	the action variable, and
	\begin{align}
		\label{eq:PolicyDef}
		\pi_\bt(\mathbf{u}|\mb{x}) = \frac{1}{ \sqrt{|\boldsymbol{\Sigma}_\mb{x}|
				(2\pi)^d}} e^{ -\frac{1}{2} (\mathbf{u}-\boldsymbol{\mu}_\mb{x})^\top
			\boldsymbol{\Sigma}_\mb{x}^{-1} (\mathbf{u}-\boldsymbol{\mu}_\mb{x})}\,,
	\end{align}
	where $\boldsymbol{\mu}_\mb{x}\in\mathbb{R}^d$ and
	$\boldsymbol{\Sigma}_\mb{x}\in\mathbb{R}^{d \times d}$ are generated by using
	an FNN denoted by $f_\bt(\mb{x})$ as
	\begin{align}
		\label{eq:FNN}
		\boldsymbol{\mu}_\mb{x} = L_{\pmb{\mu}}(f_\bt(\mb{x})), \quad
		\boldsymbol{\Sigma}_\mb{x} =  M(L_{\pmb{\Sigma}}(f_\bt(\mb{x}))),
	\end{align}
	wherein $L_{\pmb{\mu}}$ and $L_{\pmb{\Sigma}}$ are linear functions which simply map some parts of the FNN output to the $d$-dimensional vector $\pmb{\mu}_{\mb{x}}$, and other parts to the $d\times d$ dimensional matrix. In this case, $\bt$ represents the weights and biases which describe the FNN. The function $M:\real{d\times d} \mapsto \real{d\times d}$, defined as $M(\mb{A}) \ldef \mb{A}\mb{A}^\top$ ensures that the matrix $\pmb{\Sigma}_{\mb{x}}$ is always positive definite, which cannot otherwise be ensured for an arbitrary $d\times d$ matrix output from an FNN. For the rest of the paper, we refer to both policy distribution $\pi_\bt$ and
	parameter vector $\bt$ as `policy', with the exact meaning evident from the
	context. 
	
	The design of the reward function is essentially the same as that of $J_{\mcal{T}}$ in~\eqref{eq:objective}, except that it is defined over a shorter time horizon, and terms further in the future are discounted with the discount factor $\gamma$. The reward at time $t$ is given as:
	\begin{align*}
		R(\mb{x}_t,\mb{u}_t) = \textstyle\sum_{t'=t}^{t+H} &\gamma^{(t-t')} \big[  \bluen{\textstyle\sum_{i\in\mcal{N}^\text{L}} C^\text{L}_{i,t'}(\Re(s_{i,t'}))} + \\&\bluen{\textstyle\sum_{i\in\mcal{N}^\text{MT}} C^{\text{MT}}_{i,t'} (\Re(s_{i,t'}))}  \big].
	\end{align*}
	The constraints on the variables in action $\mathbf{u}_t$ are denoted by the
	vector-valued function $\mb{C} (\mb{x}_t,\mathbf{u}_t) \preceq \mb{0}$. Note that we consider only inequality constraints (`$\leq$') from load restoration in $\mcal{J}^C$, while equality constraints are embedded implicitly in the observation generation procedure, which solves~\eqref{eq:optprob} for all the unspecified variables. This allows for a simulator-free implementation of CPO; indeed,~\eqref{eq:optprob} can itself be considered a `simulator' in the current application.
	We define the \emph{reward function} and \emph{constraint function} at time $t$ as
	\begin{align*}
		\mj^R(\pi_\bt,\mb{x}_t) &\ldef \expect{\mathbf{u}_t \sim \pi_\bt}{
			R(\mb{x}_t,\mathbf{u}_t) \, | \,\mb{x}_t},\\
		\mj^{\mb{C}}(\pi_\bt,\mb{x}_t) &\ldef
		\expect{\mathbf{u}_t\sim\pi_\bt}{\mb{C}(\mb{x}_t,\mathbf{u}_t) \,| \,\mb{x}_t}.
	\end{align*}

	\begin{algorithm}[t!]
		\caption{Training load restoration policy with CPO}
		\label{alg:dataset}
		\textbf{Input:} Initial weights $\pmb{\theta}$, number of episodes $E$, batch size $B$, stale update parameter $m$, look-ahead window $H$, multiple RES and load forecasts $\{\hat{P}^\text{RES},\hat{P}^\text{L},\hat{Q}^\text{L}\}$.\\
		\textbf{Output:}  Trained policy weights $\pmb{\theta}$
		\begin{algorithmic}[1]
			\For{$e=1$ to $E$} \Comment{\texttt{episodes}}
			\State Pick a sample from $\{\hat{P}^\text{RES},\hat{P}^\text{L},\hat{Q}^\text{L}\}$ for the current episode
			\For{$t\in\mcal{T}$} \Comment{\texttt{time horizon}}
			\State Generate state $\mb{x}_{t-1}$ from $\mb{o}_{t-1}$, forecasts 
			\State Sample $\mcal{N}(\mb{0},\mb{I}^\text{id}_d)$ $B$ times as $\{ \pmb{\epsilon}^{(b)} \}_{b\in[B]}$
			\State Generate coefficients $\mb{a}_t,\mb{B}_t,\mb{c}_t$, $\mb{F}_t$ using Prop.~\ref{th:OptApprox} and Lemma~\ref{lemma:partialDerivative} \Comment{\texttt{exploration by policy}}
			\If{$\text{mod}(t,m) = 0$} \Comment{\texttt{FIM stale update}}
			\State Generate $\pmb{\Sigma}^{-1}_{\pmb{\theta}}$ using Prop.~\ref{th:OptApprox} and Lemma~\ref{lemma:partialDerivative}
			\EndIf 
			\State Update $\pmb{\theta}$ by solving~\eqref{eq:PCPOrelaxedprob}
			\EndFor
			\EndFor
			\State \textbf{return} $\pmb{\theta}$
		\end{algorithmic}
		\label{alg:CPO}
	\end{algorithm}

		The load restoration problem can be solved by determining a policy $\bt^*$
		such that for any state $\mb{x}$, $\pi_{\bt^*}$ maximizes
		$\mj^R(\pi_{\bt^*},\mb{x})$ while respecting the constraints
		$\mj^{\mb{C}}(\pi_{\bt^*},\mb{x}) \preceq \mb{0}$. Since finding a $\bt^*$ which produces the optimal action for \emph{all} possible states is an intractable problem, we instead adopt the framework of training $\bt$ episodically. Thus, a near-optimal policy $\bt^*$ can be
		found in an episodic fashion by sequentially solving the following problem:
	\begin{subequations}
		\label{eq:PCPOorigprob}
		\begin{align}
			\label{eq:pcpo1}
			\bt_{t+1} = \argmax\limits_{\bt}\,\,  &\mj^R(\pi_\bt,\mb{x}_t) \\
			%\tag{\theequation{a}}
			\label{eq:pcpo2}
			\textrm{s.t.}\quad &\mj^{\mb{C}}(\pi_\bt,\mb{x}_t) \preceq \mb{0} \\
			%\tag{\theequation{a}}
			\label{eq:KLDivCondition}
			&\kl{ \pi_\bt( \, . \,|\mb{x}_t ) }{ \pi_{\bt_t}( \, . \,|\mb{x}_t ) } \leq
			\delta,
		\end{align}
	\end{subequations}
	where $\delta>0$ is the \emph{trust region parameter} to ensure that successive
	policies do not have large variations. The final policy $\bt^*$ is then determined as $\bt^* = \lim\limits_{t\rightarrow \infty} \bt_t$. Since the original problem~\eqref{eq:optprob} is only defined for $t\in\mcal{T}$ and not for all $t\in\mathbb{N}$, we `loop back' to a different initial state $\mb{x}_0$ at time $T+1$. This idea of training multiple times over $\mcal{T}$ using different initial conditions is referred to as \emph{episodic training}.
	
	Directly solving~\eqref{eq:PCPOorigprob} is challenging due to the highly nonlinear and non-convex optimization landscape of~\eqref{eq:PCPOorigprob}, which arises due to the nonlinearity of FNNs. In order to alleviate this challenge, we replace~\eqref{eq:PCPOorigprob} with a quadratically constrained linear program (QCLP) approximation, which is well known in the literature~\cite{JA-etal:2017}.
	\begin{subequations}
		\label{eq:PCPOrelaxedprob}
		\begin{align}
			\label{eq:PCPORelax1}
			\bt_{t+1} = \argmax_{\bt}\,\, &\mb{a}_t^\top (\bt- \bt_t)\\
			\label{eq:PCPORelax2}
			\textrm{s.t.}\quad &\mb{B}_t^\top(\bt- \bt_t) + \mb{c}_t \preceq \mb{0}\\
			\label{eq:KLDivRelax}
			&(\bt- \bt_t)^\top \mb{F}_t (\bt-\bt_t) \leq \delta.
		\end{align}  
	\end{subequations}
	The above QCLP approximation is obtained by replacing the objective function~\eqref{eq:pcpo1} and constraint function~\eqref{eq:pcpo2} with their first-order Taylor approximations~\eqref{eq:PCPORelax1} and~\eqref{eq:PCPORelax2},  respectively. The KL divergence constraint~\eqref{eq:KLDivCondition} is replaced with its second-order Taylor approximation~\eqref{eq:KLDivRelax}, since its first-order approximation vanishes~\cite{OB-YIA:2013}. $\mb{F}_t$ is positive definite by construction~\cite{OB-YIA:2013}, and therefore constraint~\eqref{eq:KLDivRelax} is convex.                                                                         
	
	The challenge in adapting CPO to any given application is to find an accurate and efficient algorithm for computing $\mb{a}_t$, $\mb{B}_t$, $\mb{c}_t$, and $\mb{F}_t$ on every time step $t$. In the following proposition, we present a procedure to generate said variables on every $t$, which simply requires an expectation over a standard normal distribution, along with the partial derivative of the reward and constraint functions with respect to action. This is a significant improvement over the training procedure presented in~\cite{QZ-KD-ZW-FQ-DZ:2021}, which requires multiple matrix inversions and matrix exponential evaluations per time step.
	\newcommand{\hem}[1]{\hspace{-#1em}}
	\begin{prop}[Parameters in problem~\eqref{eq:PCPOrelaxedprob}]
		\label{th:OptApprox}
		Define the variables $\pmb{\mu}_{\bt} \ldef L_\mb{\pmb{\mu}}(f_\bt(\mb{x}_t))$ and $\pmb{\Sigma}_\bt \ldef M(L_{\pmb{\Sigma}}(f_\bt(\mb{x}_t)))$. 
		Then, we have
		\begin{gather*} 
			\mb{a}_t = \left(\expect{\pmb{\epsilon} \sim \mcal{N}
				(\mb{0},\mb{I}^\id_d)}{\frac{\partial R_t}{\partial \mathbf{u}_t} \left(
				(\pmb{\epsilon}^\top\hem{0.5} \otimes \hem{0.2}\mb{I}^\id_d) \frac{\partial \mb{v}_\bt}{\partial \bt}\hem{0.2}
				+\hem{0.2} \frac{\partial \pmb{\mu}_\bt}{\partial \bt} \right) \, \bigg\vert \, \mb{x}_t}
			\right)_{\bt=\bt_t}\\
			\mb{B}_t  = \left(\expect{\pmb{\epsilon} \sim \mcal{N}
				(\mb{0},\mb{I}^\id_d)}{\frac{\partial \mb{C}_t}{\partial \mathbf{u}_t} \left(
				(\pmb{\epsilon}^\top\hem{0.5} \otimes\hem{0.2} \mb{I}^\id_d) \frac{\partial \mb{v}_\bt}{\partial \bt}\hem{0.2}
				+\hem{0.2} \frac{\partial \pmb{\mu}_\bt}{\partial \bt} \right)\, \bigg\vert \, \mb{x}_t
			}\right)_{\bt=\bt_t}\\
			\mb{c}_t = \mj^{\mb{C}}(\pi_{\bt_t},\mb{x}_t)\\
			\mb{F}_t(i,j)\hem{0.3} =\hem{0.3} \bigg[\frac{\partial \pmb{\mu}_\bt^\top}{\partial \bt(i)} 
			\pmb{\Sigma}_\bt^{-1}\hem{0.2}
			\frac{\partial \pmb{\mu}_\bt}{\partial \bt(j)}\hem{0.3}+\hem{0.3}\frac{1}{2} \hem{0.3} \Tr \hem{0.3} \left(\hem{0.2} \pmb{\Sigma}_\bt^{-1}\hem{0.2}
			\frac{\partial\pmb{\Sigma}_\bt}{\partial \bt(i)} \pmb{\Sigma}_\bt^{-1}\hem{0.2} 
			\frac{\partial \pmb{\Sigma}_\bt}{\partial \bt(j)}\hem{0.2} \right) \bigg]_{\bt = \bt_t}
		\end{gather*}
		where $R_t \ldef R(\mb{x}_t,\mathbf{u}_t)$, $\mb{C}_t \ldef
		\mb{C}(\mb{x}_t,\mathbf{u}_t)$, $\mb{v}_\bt \ldef \text{vec} (\pmb{\Sigma}_\bt)$, and
		$\mb{F}_t(i,j)$ and $\bt(i)$ are the $\tth{(i,j)}$ and $\tth{i}$ element of
		$\mb{F}_t$ and $\bt$, respectively.
	\end{prop}
	\begin{proof}
		See Appendix~\ref{proofThm1}.
	\end{proof}
	
	In the following remark, we highlight some implementation details of the formulae presented in Proposition~\ref{th:OptApprox}.

		\begin{remark}[Implementation of Proposition~\ref{th:OptApprox}]
			Firstly, we observe that in their definitions, $\mb{a}_t$ and $\mb{B}_t$ are defined in the form of an expectation over $\pmb{\epsilon}\sim\mathcal{N}(\mb{0},\mb{I}^\id_d)$, which would result in the linear term $\pmb{\epsilon}^\top$ inside the expression being nullified. In practice, we implement the expectation through a sample average over samples of $\pmb{\epsilon}$ which are produced by a random number generator, and since we don't use infinite samples, the term $\pmb{\epsilon}^\top$ is not nullified. Secondly, we note that evaluating $\mb{F}_t$ requires inverting the matrix $\pmb{\Sigma}_\bt^{-1} \in \mathbb{R}^{d\times d}$ on every time step. As a workaround, we implement stale updates, which means that $\pmb{\Sigma}_\bt^{-1}$ is updated only once every $m$ time steps. Thirdly, note that once $\mb{x}_t$ and $\mb{u}_t$ are known, the variables in $\mb{o}_t$ can be derived by solving a constraint satisfaction problem using the constraints of~\eqref{eq:optprob}. This is equivalent to using~\eqref{eq:optprob} as an environment for the CPO agent in lieu of external simulators. \hfill$\blacksquare$
	\end{remark}
	
	The partial derivatives of $\mb{v}_\bt$, $\boldsymbol{\mu}_\bt$, and $\pmb{\Sigma}_\bt$ in Proposition~\ref{th:OptApprox} with respect to $\bt$ may be calculated via \emph{backpropagation} operation~\cite[Algorithms 6.3 and 6.4]{lecun2015deep} on the FNN, which is a standard operation for any software capable of handling FNNs. Now, it only remains to develop a procedure to evaluate $\frac{\partial R_t}{\partial \mb{u}_t}$ and $\frac{\partial \mb{C}_t}{\partial \mb{u}_t}$. The procedure we develop is unique to the DistFlow equations, and is similar for both the terms, and therefore we only demonstrate the evaluation of $\frac{\partial R_t}{\partial \mb{u}_t}$. 
	%We demonstrate the procedure only for single phase MGs, with Appendix~\ref{app:mphase} delving into its extension for multiphase systems. 
	
	For generic variables $a$ and $b$, let the expressions $a\in R_t$ and $b\in\mb{u}_t$ imply that $a$ makes an appearance in the closed form of $R_t$, and $b$ is a variable in $\mb{u}_t$. The evaluation of $\frac{\partial R_t}{\partial \mb{u}_t}$ therefore boils down to the efficient evaluation of $\frac{\partial a}{\partial b}$. To this end, we collect the equations which couple variables in $\mb{x}_t$, $\mb{u}_t$, and $\mb{o}_t$ over the time horizon $\{ t,\cdots,t+H-1 \}$, and evaluate their total differentials as follows:
	\begin{subequations}
		\label{eq:totalDeriv}
		\begin{gather}
			\label{eq:TD1}
			ds_{i,t} = \sum_{j \rightarrow k} dS_{jk,t} - \sum_{i \rightarrow j} (dS_{ij,t}-z_{ij}dl_{ij,t}),\\
			\label{eq:TD3}
			dv_{j,t} = dv_{i,t} - 2\Re(\bar{z}_{ij}dS_{ij,t}) + |z_{ij}|^2dl_{ij,t},\\
			\label{eq:TD4}
			dl_{ij,t}v_{i,t} + l_{ij,t}dv_{j,t} = 2\Re(\bar{S}_{ij,t}dS_{ij,t}),\\
			\label{eq:TD5a}
			P^\text{dis}_{i,t} dP^\text{ch}_{i,t} + dP^\text{dis}_{i,t} P^\text{ch}_{i,t} = 0,\\
			\label{eq:TD5}
			\Re(ds_{i,t}) =dP^{\text{dis}}_{i,t} -  dP^{\text{ch}}_{i,t}, \\
			\label{eq:TD6}
			d\mcal{S}_{i,t+1} = d\mcal{S}_{i,t} + (\eta^\text{ch}_i\Delta_t)
			dP^\text{ch}_{i,t} - \left( \frac{\Delta_t}{\eta^\text{dis}_i} \right)
			dP^\text{dis}_{i,t},\\
			\label{eq:TD7}
			d\zeta_{i,t+1} = d\zeta_{i,t} - \tau_i \Re(ds_{i,t}),\\
			\label{eq:TD8a}
			\Re(ds_{i,t}) = -d\rho_{i,t} \hat{P}^\text{L}_{i,t}, \;\; \Im(ds_{i,t}) = -d\rho_{i,t} \hat{Q}^\text{L}_{i,t}\\
			\label{eq:TD8b}
			\Re(ds_{i,t}) = -d\kappa_{i,t} \hat{P}^\text{RES}_{i,t},\\
			\label{eq:TD8}
			d\omega_t = -k_P\Re(ds_{i,t}),\\
			\label{eq:TD9}
			\left( \sqrt{v_{i,t}} \right)^{-1} dv_{i,t} = -2k_Q\Re(ds_{i,t}),
		\end{gather}
	\end{subequations}
	where \eqref{eq:TD1} holds for all
	$j\in\mcal{N}$,~\eqref{eq:TD3}--\eqref{eq:TD4} hold for all
	$i\rightarrow j$,~\eqref{eq:TD5a}--\eqref{eq:TD6} hold for all
	$i\in\mcal{N}^\text{ESS}$,\eqref{eq:TD7} holds for all
	$i\in\mcal{N}^\text{MT}$,~\eqref{eq:TD8a} holds for all $i\in\mcal{N}^\text{L}$,~\eqref{eq:TD8b} holds for all $i\in\mcal{N}^\text{RES}$, and~\eqref{eq:TD8}-\eqref{eq:TD9} hold for all $i\in\mcal{N}^\text{droop}$. Note that
	\eqref{eq:totalDeriv} is a homogeneous system of linear equations in the
	differentials, which can be used to numerically calculate the required partial
	derivatives via the following result.
	\begin{lemma}\label{lemma:partialDerivative}
		The partial derivative $\frac{\partial a}{\partial b}$ for $a\in R_t$ and $b\in \mb{u}_t$ may be evaluated by setting $db'=0$ for all $b'\in\mb{u}_t$ such that $b\neq b'$, and solving~\eqref{eq:totalDeriv} for $da$ and $db$. A solution exists if number of loads exceeds non-loads, in which case $\frac{\partial a}{\partial b} = \frac{da}{db}$.
	\end{lemma}
	\begin{proof}
		See Appendix~\ref{proofLemma3}.
	\end{proof}
	All the steps involved in CPO training are summarized in Algorithm~\ref{alg:CPO}.
	
	\subsubsection*{MG implementation}
	After we train the FNN for a large number of time steps $t$ in an episodic fashion, the trained FNN may be used as the MGC. If deterministic policies are desired, then the mean $\pmb{\mu}_{\mb{x}}$ may be used while the covariance matrix $\pmb{\Sigma}_{\mb{x}}$ may be nullified. On the other hand, if stochastic policies are desired, both $\pmb{\mu}_{\mb{x}}$ and $\pmb{\Sigma}_{\mb{x}}$ may be retained.
	
	\section{Simulation Results}\label{sec:simulation}
	\begin{figure}[!tb]
		\centering
		\includegraphics[width=0.75\linewidth]{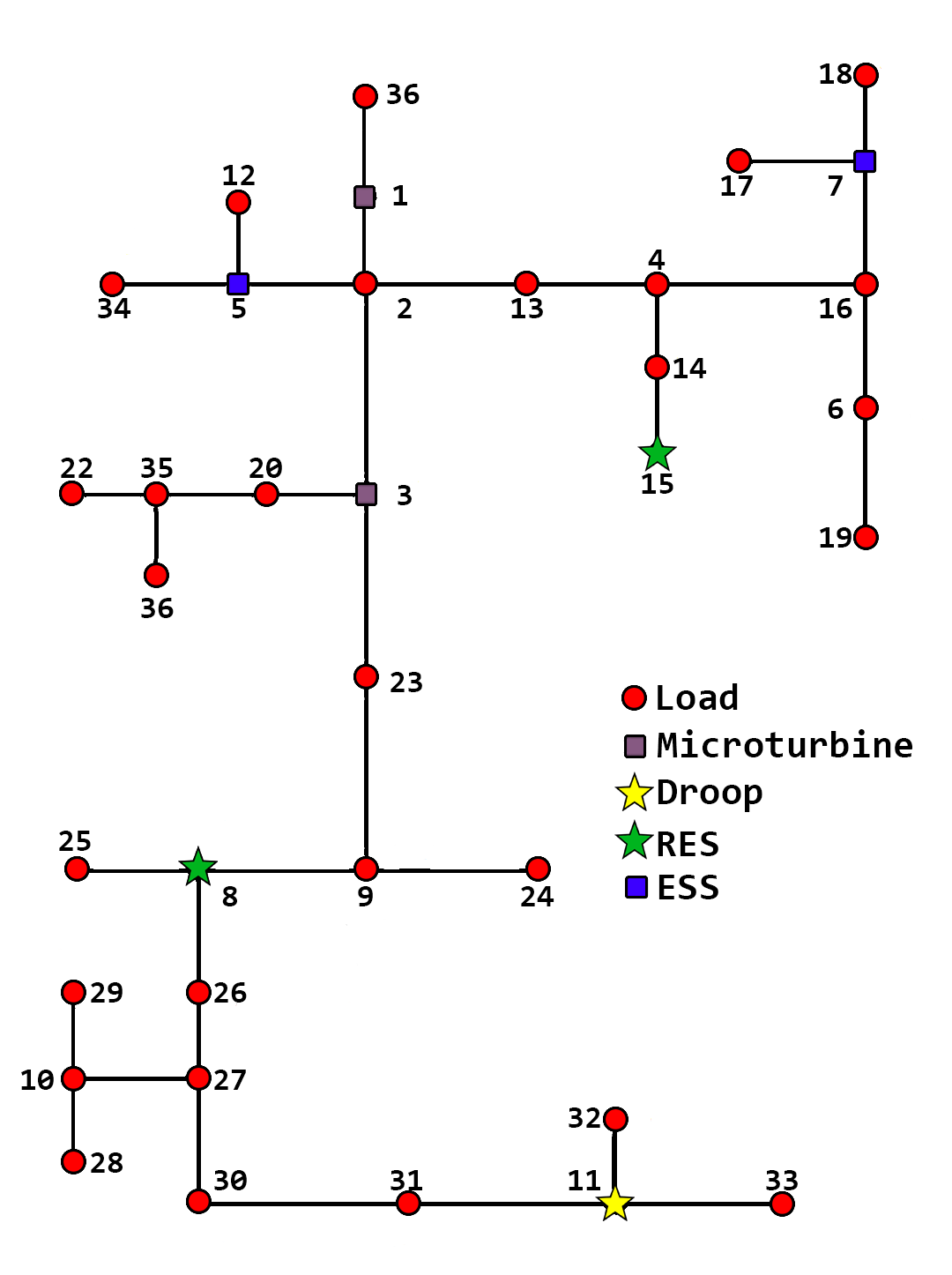}
		\caption{{A 36-bus MG that is adapted from the IEEE 37-bus distribution feeder.}}
		\label{fig:36}
	\end{figure}
	\begin{figure*}
		\centering
		\begin{subfigure}[b]{0.3\textwidth}
			\centering
			\includegraphics[width=\textwidth]{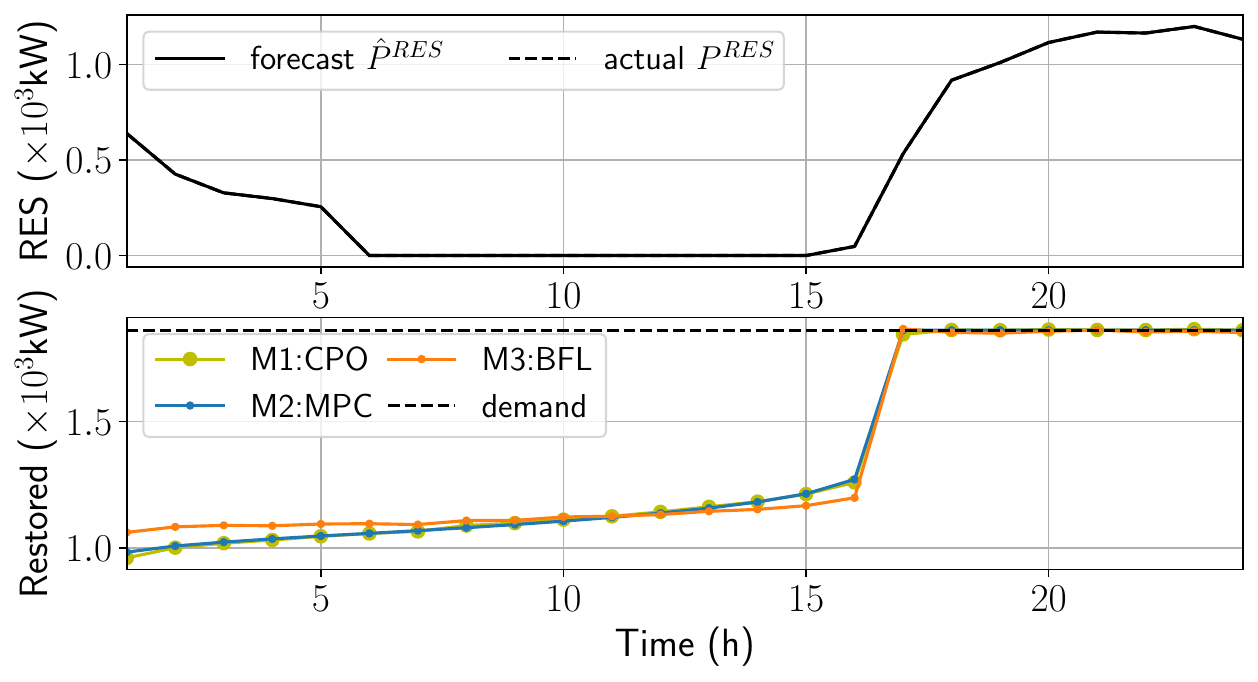}
			\caption{}
			\label{fig:33uncertainty0}
		\end{subfigure}
		\hfill
		\begin{subfigure}[b]{0.3\textwidth}
			\centering
			\includegraphics[width=\textwidth]{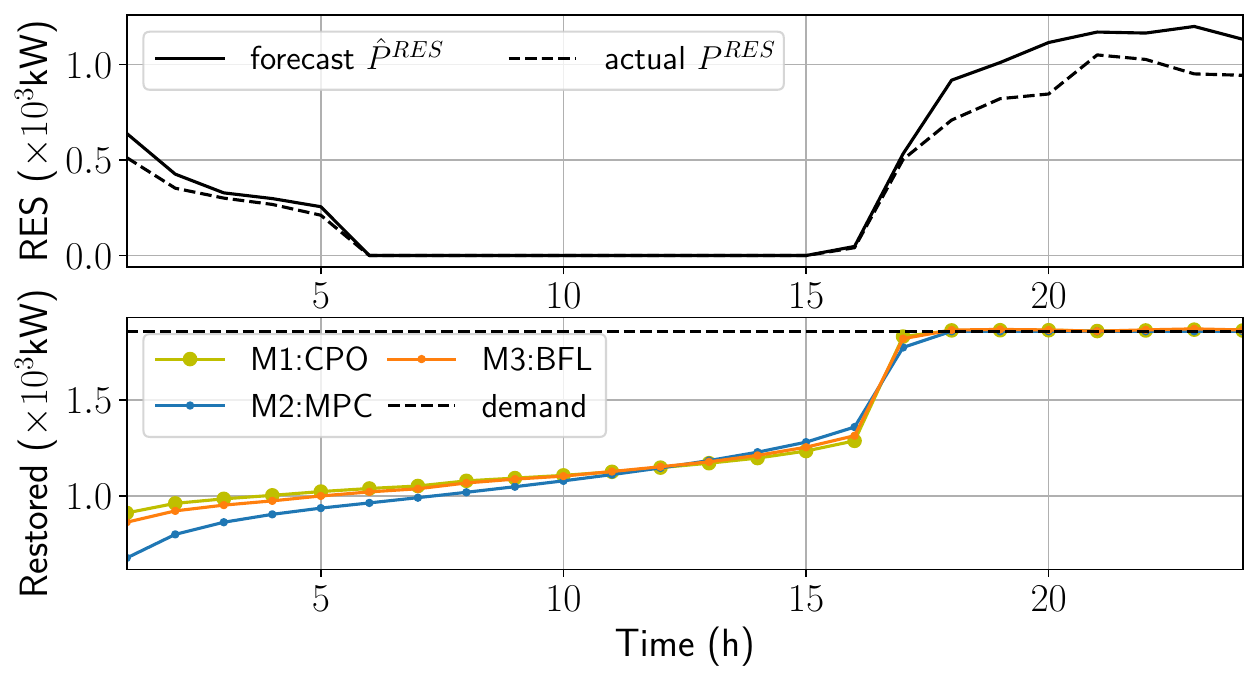}
			\caption{}
			\label{fig:33uncertainty1}
		\end{subfigure}
		\hfill
		\begin{subfigure}[b]{0.3\textwidth}
			\centering
			\includegraphics[width=\textwidth]{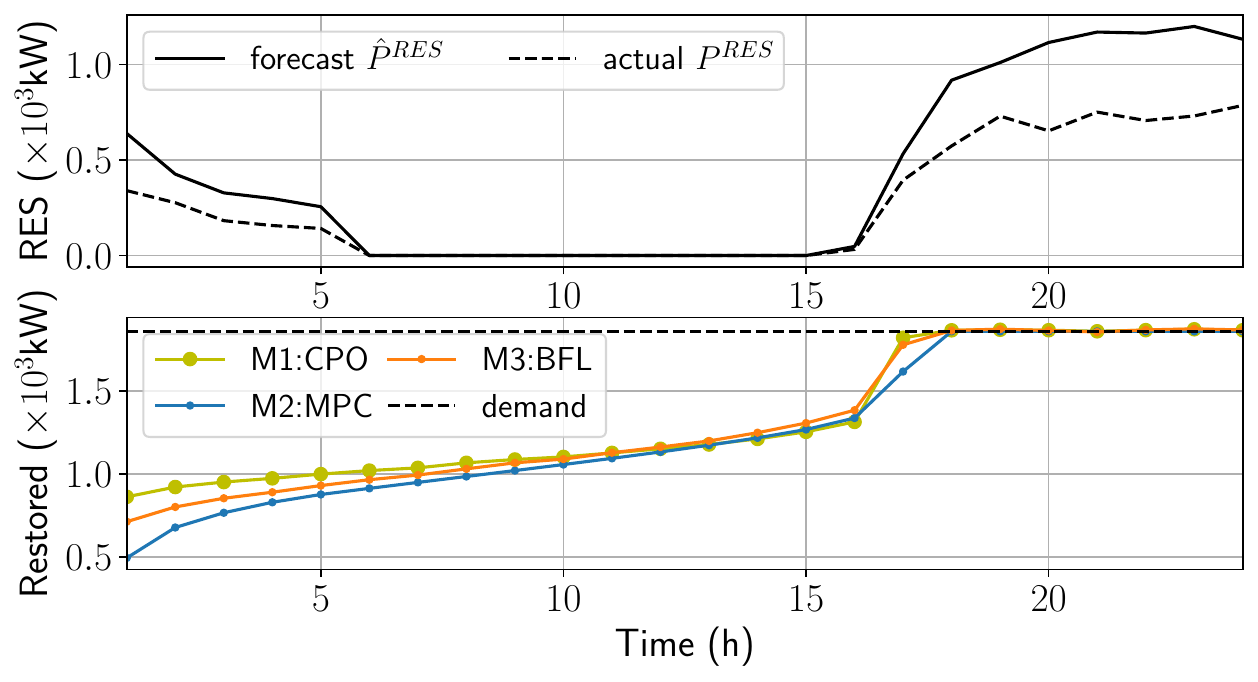}
			\caption{}
			\label{fig:33uncertainty2}
		\end{subfigure}	
		\newline
		\label{fig:three graphs}
		\begin{subfigure}[b]{0.3\textwidth}
			\centering
			\includegraphics[width=\textwidth]{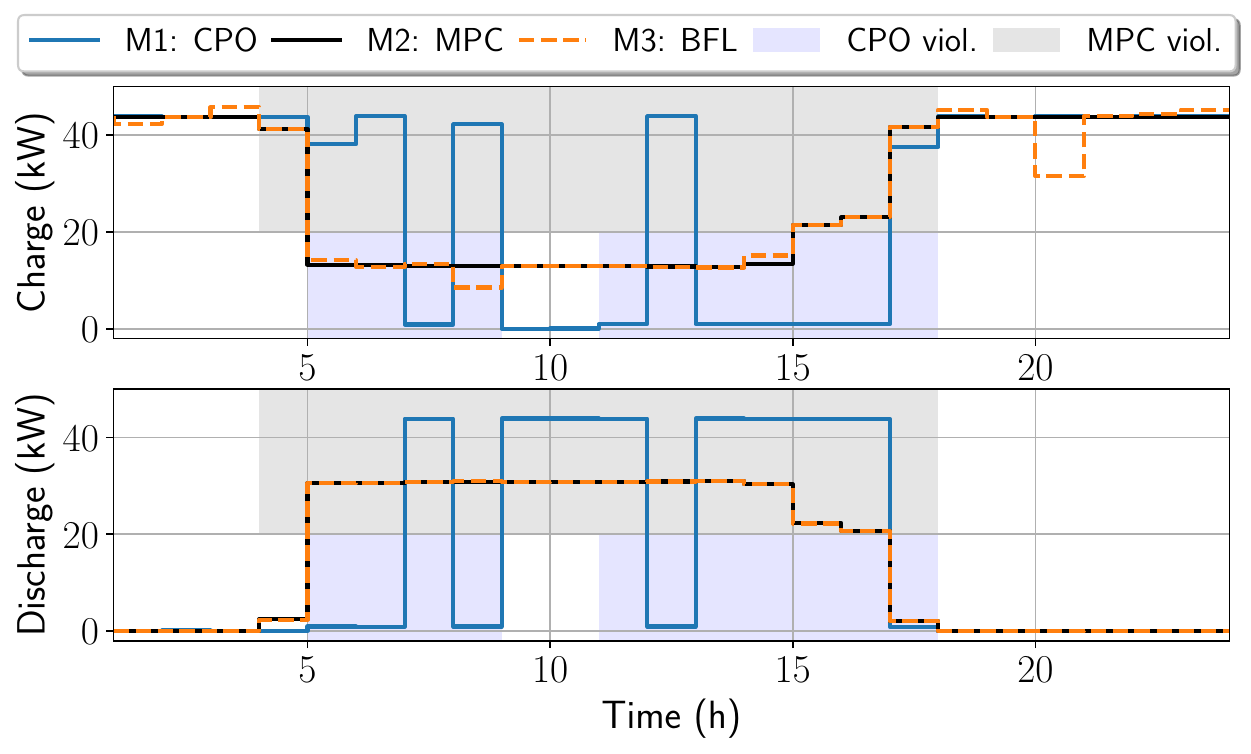}
			\caption{}
			\label{fig:chargedis33}
		\end{subfigure}
		\hfill
		\begin{subfigure}[b]{0.3\textwidth}
			\centering
			\includegraphics[width=\textwidth]{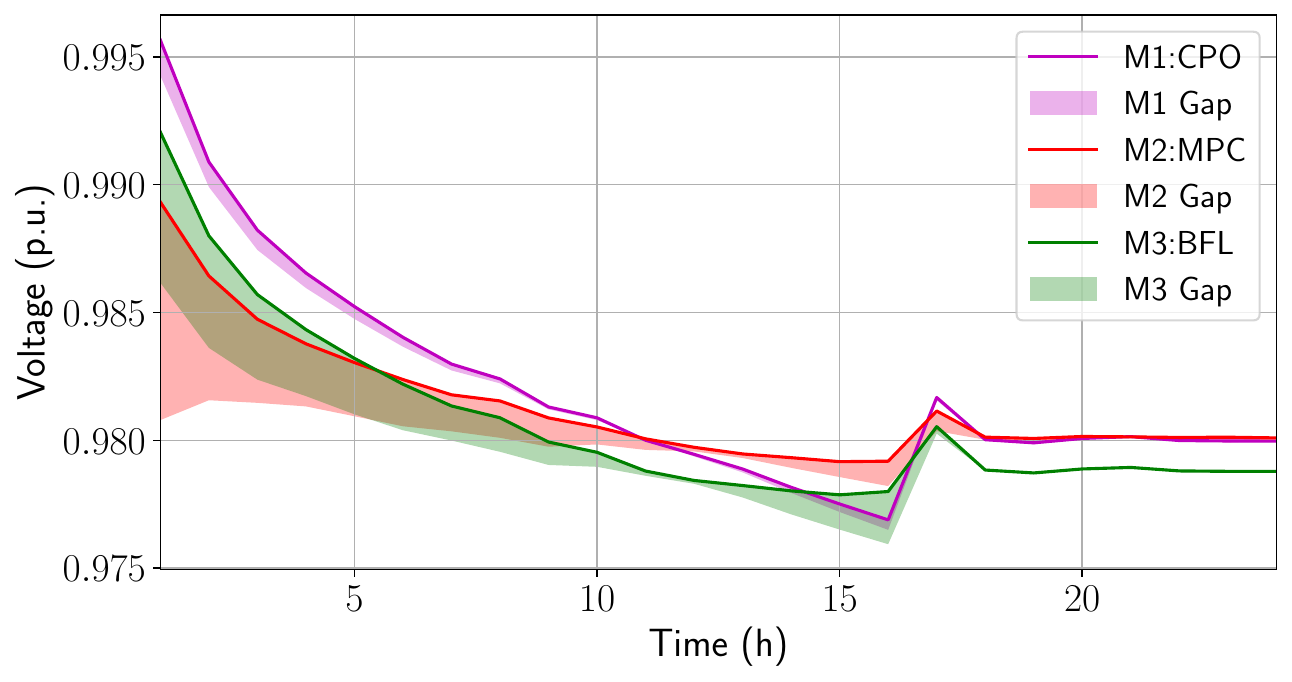}
			\caption{}
			\label{fig:voltagedroop}
		\end{subfigure}
		\hfill
		\begin{subfigure}[b]{0.3\textwidth}
			\centering
			\includegraphics[width=\textwidth]{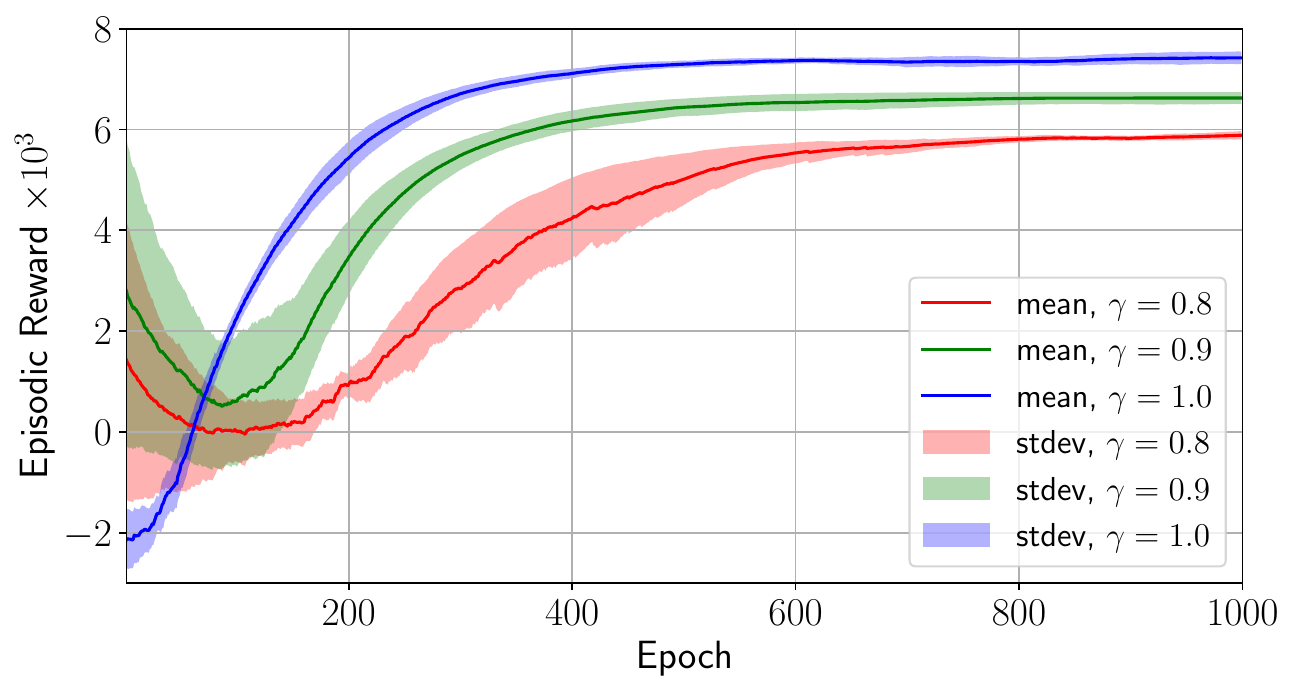}
			\caption{}
			\label{fig:reward33}
		\end{subfigure}
		\caption{Simulation results for the 36-bus MG.}
		\label{tab:33bus}
	\end{figure*}
	\begin{figure*}[!tb]
		\centering
		\begin{subfigure}[b]{0.3\textwidth}
			\centering
			\includegraphics[width=\textwidth]{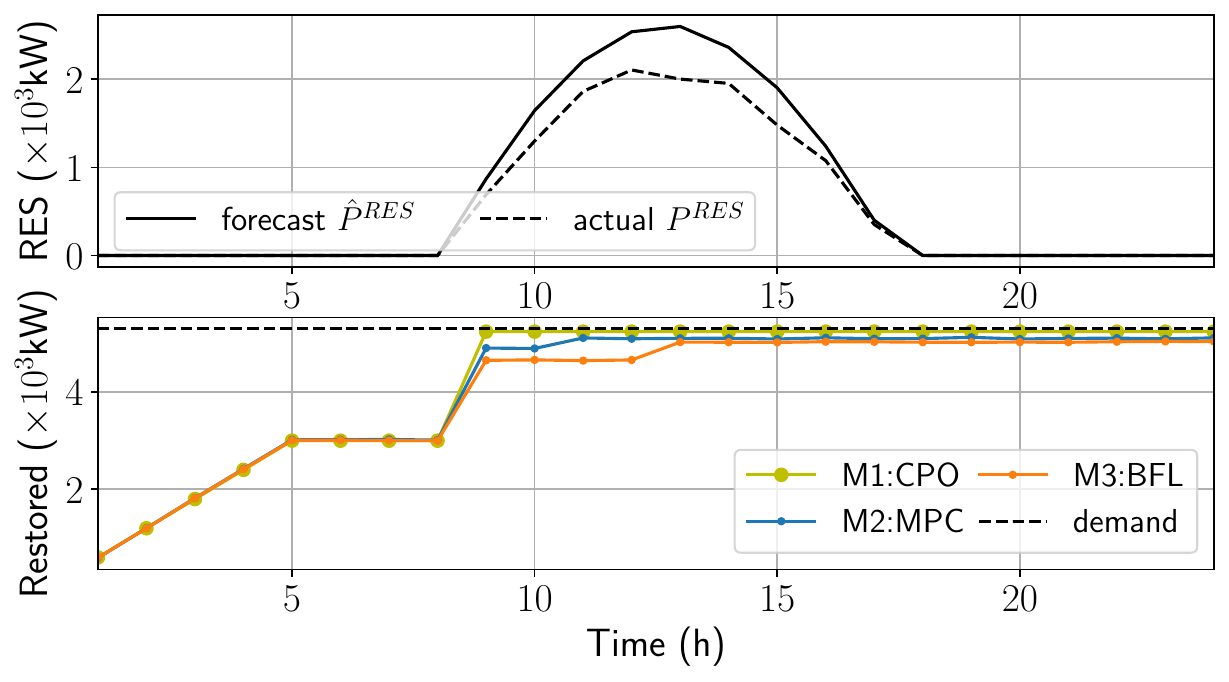}
			\caption{}
			\label{fig:141restore}
		\end{subfigure}
		\hfill
		\begin{subfigure}[b]{0.3\textwidth}
			\centering
			\includegraphics[width=\textwidth]{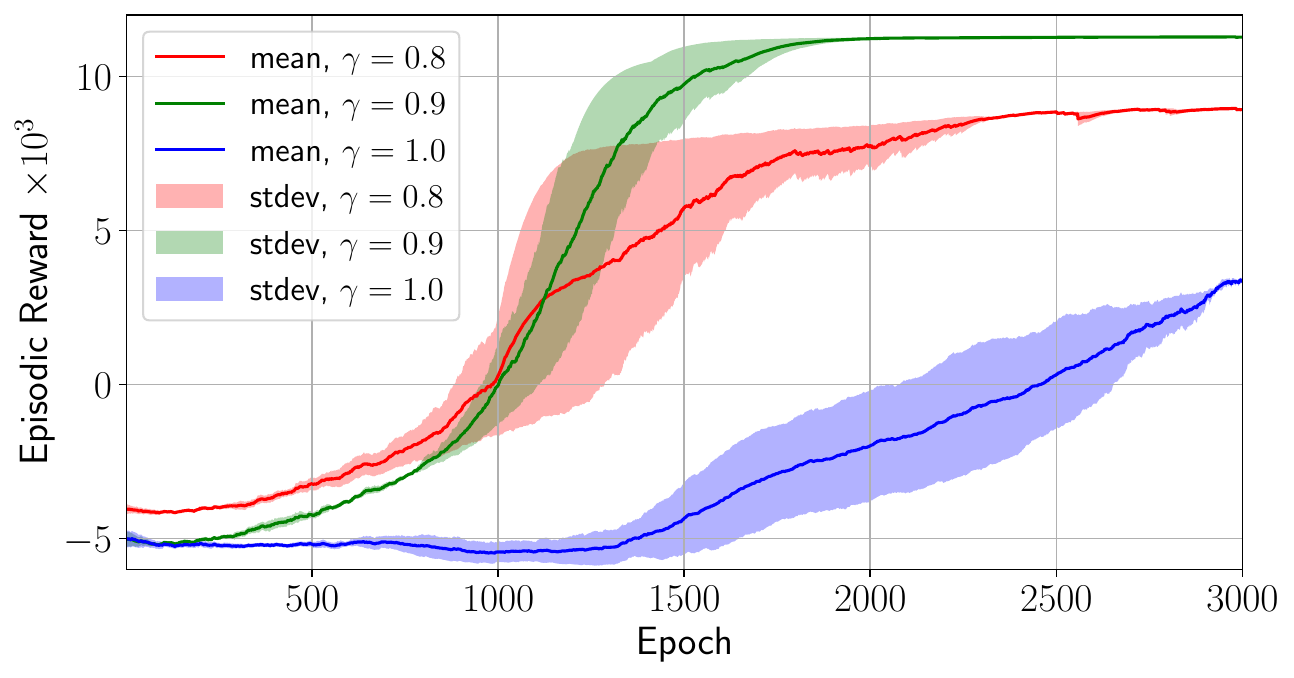}
			\caption{}
			\label{fig:reward141}
		\end{subfigure}
		\hfill
		\begin{subfigure}[b]{0.3\textwidth}
			\centering
			\includegraphics[width=\textwidth]{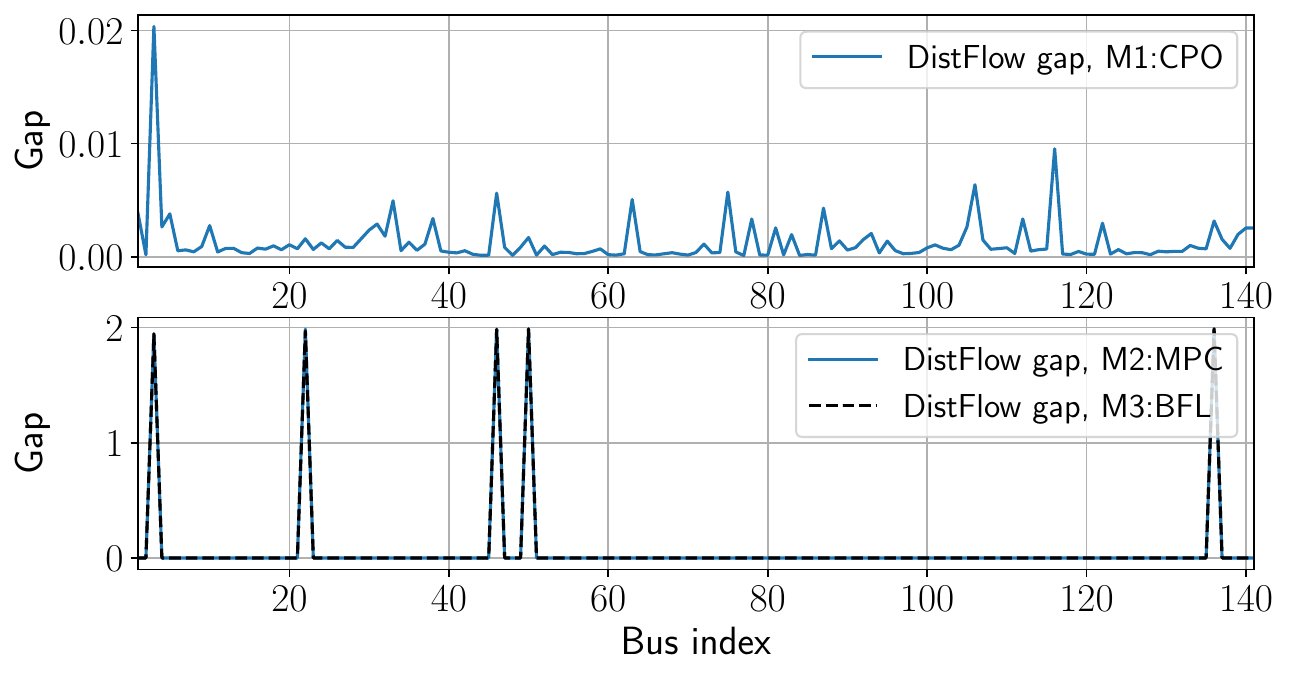}
			\caption{}
			\label{fig:ldfdualitygap}
		\end{subfigure}	
		\caption{Simulation results for the 141-bus MG.}
	\end{figure*}
	
	\begin{table}[!bt]
		\centering
		\caption{System Parameters.}
		\label{tab1}
		\begin{tabular}{|c|c|}
			\hline 
			\textbf{System Parameter} & \textbf{Value}\\
			\hline\hline
			$T (\text{hours})$ & 24\\
			\hline
			$(\underline{v} (\text{p.u.}),\bar{v} (\text{p.u.}))$ & $(0.95, 1.05)$\\
			\hline
			$(\underline{P}^\text{MT}(\text{kW}),\overline{P}^\text{MT}(\text{kW}),\tau (\text{kW}^{-1}))$
			& $(0, 250, 0.9)$\\
			\hline
			$(E_0,\underline{P}^\text{MT}_\text{rd}(\text{kW}),\bar{P}^\text{MT}_\text{ru}(\text{kW}))$&$(930, -30, 25)$\\
			\hline
			$(\bar{P}^\text{ch}(\text{kW}),\bar{P}^\text{dis}(\text{kW}),\eta^\text{ch},\eta^\text{dis})$
			& $(25,25,0.9,0.9)$\\
			\hline
			$\underline{\mcal{S}}(\text{kW}),\bar{\mcal{S}}(\text{kW}),\mcal{S}^\text{init}$ & $(30,900,665)$\\
			\hline
%			$(\mu_\text{RES},\sigma_\text{RES},\epsilon,\Delta_t)$ &
%			$(67.3,5,0.02,1\text{ hr})$\\
%			\hline
			$(\bar{Q}^\text{MT}(\text{kVar}),\bar{Q}^\text{ESS}(\text{kVar}),\bar{Q}^\text{RES}(\text{kVar}))$ &
			$(35,35,70)$\\
			\hline
		\end{tabular}
	\end{table}
	
	\begin{table}[!tb]
		\centering
		\caption{CPO Parameters.}
		\label{tab2}
		\begin{tabular}{|c|c|c|}
			\hline 
			\textbf{CPO Parameter} & \textbf{36-bus MG} & \textbf{141-bus MG}\\
			\hline\hline
			FNN $f_\bt$ layer sizes & $(344,100,75,440)$ & $(1403,500,250,1630)$\\
			\hline
			Activation function & $\tanh$ & $\tanh$\\
			\hline
			Trust region $\delta$ & 0.001 & 0.008\\
			\hline 
			$\pmb{\epsilon}$ samples per time step & $32$ & $64$\\
			\hline
			Stale update param. $m$ & 5 & 5\\
			\hline
		\end{tabular}
	\end{table}
		In this section, we use simulations to validate the performance of the proposed MPC and CPO approaches. In order to model the MG, we use a modified IEEE 37-bus test feeder~\cite{ieee37}, which prescribes network topology and power injection data. The 37-bus system is modified to a 36-bus system by deleting a bus interfaced with the network through a transformer, thereby maintaining the same voltage levels across the MG. We also consider an MG based on a larger 141-bus radial feeder derived from \texttt{case141} in MATPOWER~\cite{RDZ-etal:2011}. 
		
		All simulations were carried out in Python on a PC with an Intel Core i7 CPU, NVIDIA 1060Ti GPU, and 32GB of RAM. The Gurobi solver~\cite{gurobi}, interfaced with python through CVXPY~\cite{cvx}, was used to find solutions for the MPC problem as well as solving~\eqref{eq:PCPOrelaxedprob}. All FNN operations, as well as gradient calculations for Lemma~\ref{lemma:partialDerivative} are calculated with PyTorch, and the automatic differentiation engine PyTorch Autograd.
		
		We tested three algorithms which can solve~\eqref{eq:optprob} as follows:
		\begin{itemize}
			\item \textbf{M1}: This method uses a stochastic CPO policy trained according to~\eqref{eq:PCPORelax1} for 1000 episodes for the 36-bus MG and 3000 episodes for the 141-bus MG. We use $H=5$ and $\gamma={0.8,0.9,1.0}$, and retain the value of $\gamma$ with the best performance. The parameters for the FNN used as the CPO policy, as well as other parameters involved in CPO, are displayed in Table~\ref{tab2}. To relieve computational burden, we choose $\pmb{\Sigma}_\bt$ to be a diagonal matrix.
			\item \textbf{M2}: This method uses MPC to solve the proposed convex relaxation. We use $H=5$ and the convex relaxations proposed in Section~\ref{sec:MPC}.
			\item \textbf{M3}: We refer to this method as \emph{brute-force learning} (BFL). It involves generating state-action pairs $(\mb{x}_k,\mb{u}_k)$, wherein the optimal actions are generated via M2. The data are generated for various renewable forecasts, similar to Algorithm~\ref{alg:CPO}. Then, we train an FNN on the input-output pairs. This method evaluates the performance of deep architectures which are trained in a supervised fashion, and are not privy to constraint-respecting training as in CPO.
			% Similar approaches have recently gained traction for solving optimal power flow problems~\cite{Fioretto_Mak_Van_Hentenryck_2020,WH-XP-MC-SHL:2022}.
		\end{itemize}
		
		\subsubsection*{36-bus MG}
		 We first consider simulation results for the 36-bus MG. The parameters for various elements in the MG are listed in Table~\ref{tab1}, and a one-line diagram of the MG is presented in Figure~\ref{fig:36}. We model the droop bus as an additional MT with similar parameters. The cost functions in~\eqref{eq:objective} are chosen as $C^\text{L}_{i,t}(\Re(s_{i,t})) = -\Re(s_{i,t})$ (since load demands are negative) and $C^\text{MT}_{i,t}(\Re(s_{i,t})) = -0.75\Re(s_{i,t})$. The RES outputs are modeled as having a particular shape as shown in Figures~\ref{fig:33uncertainty0}-\ref{fig:33uncertainty2}, and during training, they are perturbed uniformly by a factor of $[0.75,1.25]$ on each time step.
		The loads are chosen to be constant for the training in order to better demonstrate load restoration performance, and its values are derived from the case data. 
		
		The training curves for different values of $\gamma$ are presented in Figure~\ref{fig:reward33}. In the figure, the solid line shows the mean reward collected, while the shaded region indicates the variance of rewards over multiple training runs. In total, 500 training runs of Algorithm~\ref{alg:CPO} were carried out and the best policy was used to generate final results and plots. From the figure, it can be seen that setting $\gamma=1$ leads to highest reward collection and also the lowest variance during training. Thus, $\gamma=1$ is used for the remaining experiments with the 36-bus MG. Next, we compare the load restoration process with uncertain values of $\hat{P}^\text{RES}$. In order to compare the load restoration performance of the three competing methods, we consider three scenarios as shown in Figures~\ref{fig:33uncertainty0}-\ref{fig:33uncertainty2}. In the first scenario with perfect forecast, the performance of CPO is slightly worse than MPC and BFL, which demonstrates that under the availability of perfect information, MPC solutions can be of higher quality than CPO. However, in scenarios where the actual RES output is lower than forecasts, as shown in Figures~\ref{fig:33uncertainty1}-\ref{fig:33uncertainty2}, CPO shows better performance in terms of initial load pickup, as well as achieving full load restoration. This is due to the capability of CPO to learn from experience, and therefore during training it learns the best schedule for load restoration even under imperfect information. As opposed to this, MPC is restricted to only using forecast inputs for the current time step to $H$ time steps in the future, which poses a disadvantage in uncertain systems.
		
		Next, we consider two relaxations proposed in Section~\ref{sec:MPC}. Figure~\ref{fig:chargedis33} shows the charge and discharge performance for the ESS. Recall that for MPC, we use the relaxation proposed in Lemma~\ref{lemma:ConvexHull}, and the same is also inherited BFL during its training. The gray shaded areas represent the time steps when MPC does not respect CC, while the blue shaded areas do the same for CPO. From here, it can be seen that CPO creates solutions that adhere better to CC than MPC. Furthermore, even in the time steps when CPO violated CC, the magnitude of its violation is much lower than the solution produced by MPC and BFL. Therefore, we conclude that it is possible to implement the charge/discharge schedule produced by CPO directly to the MG, while MPC and BFL schedules require the \emph{post-hoc} modification as discussed in Section~\ref{sec:MPC}.
		
		Finally, we consider the relaxation for voltage droop constraint presented in Lemma~\ref{lemma:DroopBus}. It is possible to calculate the gap between the voltages derived as a result of the relaxed constraint, and the actual voltage of the unrelaxed constraint by evaluating~\eqref{eq:droop_reactive} using reactive power injections. From here, it can be seen that MPC and BFL incur large gaps between the relaxation and the actual voltages, especially in the initial time steps. This shows that the convex relaxation proposed in Lemma~\ref{lemma:DroopBus} may not be tight in practice. However, the performance of CPO is much better in terms of the gap, thereby demonstrating that CPO is a better approach to satisfy the nonlinear droop constraints than relaxing them to a convex inequality constraint. However, further studies are required to establish conditions under which the proposed relaxation is tight, and holds exactly for all time steps.
		
		\subsubsection*{141-bus MG}
		We now consider the simulation results for the 141-bus MG. The load demands are chosen from the case data, and multiple generation sources are incorporated into the MG. We place 7 MTs at buses $\{3,51,56,77,91,123,138\}$, 3 RESs at buses $\{12,43,88\}$, 6 ESS at buses $\{16,41,77,83,124,139\}$, and buses 56 and 91 operate under the principle of droop control. As in the 36-bus MG, the total load demands to be restored are constant and shown in Figure~\ref{fig:141restore}, while the RES forecasts (with mispredictions) are also displayed in the same figure.
		
		The training process of the CPO agent, as shown in Figure~\ref{fig:reward141} shows that the best choice for parameter $\gamma$ is $\gamma=0.9$. Unlike the 36-bus MG, choosing $\gamma=1$ leads to a very unstable training trajectory, wherein the rewards remain low throughout training. Thus, we choose $\gamma=0.9$ for our experiments. The load restoration process under uncertain RES forecasts, as shown in Figure~\ref{fig:141restore}, shows that CPO performs slightly better than MPC and BFL, both in terms of load restored on any given time step, as well as reaching full load restoration. Furthermore, we compare the performance of CPO in satisfying the nonconvex DistFlow equation~\eqref{eq:PFcons3}. Recall that MPC and BFL use the relaxed second-order cone version of this constraint discussed in Section~\ref{sec:MPC}. In Figure~\ref{fig:ldfdualitygap}, we calculate the gap resulting from this constraint not binding for all three methods. It can be seen that the gap for CPO is an order of magnitude lower (in terms of maximum value) than MPC or BFL. On the other hand, MPC and BFL enjoy zero gap on most buses, but an extremely high value of the gap on certain buses. Thus, the solutions produced by these two methods enjoy lower exactness than the one produced by CPO.
		
		Finally, we discuss the time taken for both training and online runtime for all three methods, which are presented in Table~\ref{tab3}. The training time of CPO involves all steps presented in Algorithm~\ref{alg:CPO}, while the training time of BFL includes time taken for the generation of data samples as well as training the FNN on these samples. While the training time for CPO is significantly longer than BFL, it also produces small runtimes. On the other hand, since MPC has to solve multiple optimization problems per time horizon, it results in the longest runtime. BFL has a smaller runtime than CPO due to a simpler FNN structure which does not take variance into account. From here, we see that CPO is competitive with respect to MPC in terms of runtime but this comes with a tradeoff of a far longer training time, which is absent in case of MPC.
		\begin{table}[t!]
		\centering
  		\caption{Time taken by all three methods.}
		\begin{tabular}{|c|c|c|c|c|}
			\hline
			\multirow{2}{*}{\textbf{Method}} & \multicolumn{2}{c|}{\textbf{36-bus MG time}} & \multicolumn{2}{c|}{\textbf{141-bus MG time}}\\
			\cline{2-5}
			&\textbf{Training} & \textbf{Runtime} & \textbf{Training} & \textbf{Runtime}\\
			\hline
			CPO & 4310s & 0.51s & 33780s & 0.76s\\
			MPC & - & 2.76s & - & 7.59s\\
			BFL & 345s & 0.38s & 1215s & 0.62s\\
			\hline
		\end{tabular}
		\label{tab3}
		\end{table}

	\section{Conclusion}\label{sec:summary}
	In this paper we considered the load restoration problem for an islanded MG,
	which contains sources of distributed power generation such as RES and MTs, as
	well as sources of energy storage, such as ESS. Two approaches to
	find a solution to the problem which can be implemented in the MGC were studied. We
	considered MPC, and proposed a convex relaxation of the load restoration problem
	which can be efficiently solved. We also developed CPO that finds an optimal
	policy through episodic training. Then, we compared the performance of
	MPC and CPO on 36-bus and 141-bus MGs. An important direction of future extension is to consider topology-switching MGs with discrete decision variables, and implement policies for the same using CPO. Other directions of investigation involve reducing CPO training time through more efficient training strategies, and modeling of power sharing strategies in highly unbalanced multiphase MGs.
	
	\appendix
	
	\subsection{Proof of Lemma~\ref{lemma:ConvexHull}}
	\label{proofLemma2}
	We will demonstrate that the provided second-order cone can be derived from the equation $\sqrt{v_{i,t}} = \sqrt{v_i^*} - k_Q(\Im(s_{i,t})-Q_i^*)$ through a sequence of relaxations.
	We denote relaxations through the $\stackrel{\text{relax}}{\rightarrow}$ symbol.
	\begin{align*}
		&\sqrt{v_{i,t}} = \sqrt{v_i^*} - k_Q(\Im(s_{i,t})-Q_i^*)\\
		\therefore  \; & v_{i,t} = \left( \sqrt{v_i^*} - k_Q(\Im(s_{i,t})-Q_i^*)\right)^2\\
		\therefore \;&\left(v_{i,t}\right)^2 + v_{i,t} + \frac{1}{4} = \left( \sqrt{v_i^*} - k_Q(\Im(s_{i,t})-Q_i^*)\right)^2 +\\
		&\hspace{1.25in} \left(v_{i,t}\right)^2  + \frac{1}{4}\\
		\therefore \;& \left( v_{i,t} + \frac{1}{2} \right)^2 = \left\lVert \bm{ \sqrt{v_i^*} - k_Q(\Im(s_{i,t})-Q_i^*) & v_{i,t} & \frac{1}{2} } \right\rVert_2^2\\
		\therefore \;& v_{i,t} + \frac{1}{2}  = \left\lVert \bm{ \sqrt{v_i^*} - k_Q(\Im(s_{i,t})-Q_i^*) & v_{i,t} & \frac{1}{2} } \right\rVert_2\\
		\stackrel{\text{relax}}{\rightarrow} \; & v_{i,t} + \frac{1}{2}  \geq \left\lVert \bm{ \sqrt{v_i^*} - k_Q(\Im(s_{i,t})-Q_i^*) & v_{i,t} & \frac{1}{2} } \right\rVert_2. 
	\end{align*}
	
	\subsection{Proof of Lemma~\ref{lemma:DroopBus}}
	\label{proofLemma1}
	Note that the points $(0,0)$, $(0,\bar{P}^\text{dis}_i)$, and
	$(\bar{P}_i^\text{ch},0)$ are contained in $\mcal{P}_i$, and therefore any
	convex set which contains $\mcal{P}_i$ should contain
	$\operatorname{conv}\cbrace{(0,0),(0,\bar{P}_i^\text{dis}),(\bar{P}^\text{ch}_i,0)}$,
	which is exactly $\mcal{P}^\text{conv}_i$. Thus, $\mcal{P}^\text{conv}_i
	\subseteq \operatorname{conv}(\mcal{P}_i)$. On the other hand,
	$\mcal{P}_i^\text{conv}$ is convex and contains the set $\mcal{P}_i$, and
	therefore $\operatorname{conv}(\mcal{P}_i) \subseteq \mcal{P}^\text{conv}_i$. It
	follows that $\operatorname{conv}(\mcal{P}_i) = \mcal{P}^\text{conv}_i$.
	
	\subsection{Proof of Theorem~\ref{th:OptApprox}}
	\label{proofThm1}
	The main idea of the proof is to replace the reward function, constraint
	function, and KL-divergence terms in~\eqref{eq:PCPOorigprob} with their
	respective Taylor-series approximations around $\bt_t$. The first-order
	approximation of the objective function is given as
	\begin{align*}
		\mj^R(\pi_\bt,\mb{x}_t) \approx  \mj^R(\pi_{\bt_t},\mb{x}_t) + \nabla_{\bt}
		\mj^R(\pi_\bt,\mb{x}_t) \big\vert_{\bt = \bt_t}(\bt-\bt_t),
	\end{align*}
	and by comparing the above approximation with~\eqref{eq:PCPOrelaxedprob}, we
	see that $\mb{a}_t^\top = \nabla_{\bt}\mj^R(\pi_\bt,\mb{x}_t) \big\vert_{\bt =
		\bt_t}$. In order to compute $\mb{a}_t$, we need a closed-form of the gradient
	\begin{align*}
		\nabla_\bt \mj^R(\pi_\bt, \mb{x}_t) = \nabla_{\bt}
		\expect{\mathbf{u}_t\sim\pi_{\bt}}{R(\mb{x}_t,\mathbf{u}_t) \, \big| \,
			\mb{x}_t},
	\end{align*}
	which is difficult to evaluate since the gradient is with respect to $\bt$
	which parameterizes the distribution over which the expectation is being taken.
	To alleviate this difficulty, we use the \emph{reparametrization trick}. For a
	standard normal vector $\pmb{\epsilon} \sim\mcal{N}(\mb{0},\mb{I}^\id_d)$, it holds
	that
	\begin{align*}
		\pmb{\Sigma}_\bt \pmb{\epsilon} + \pmb{\mu}_\bt \sim
		\mathcal{N}(\pmb{\mu}_\bt,\pmb{\Sigma}_\bt\pmb{\Sigma}_\bt^\top).
	\end{align*}
	Therefore, letting $\mathbf{u}_t =\pmb{\Sigma}_\bt \pmb{\epsilon} + \pmb{\mu}_\bt$
	is equivalent to defining $\mathbf{u}_t$ as a Gaussian random vector with mean
	and variance given in~\eqref{eq:FNN}.
	Thus, we have
	\begin{align}
		\notag
		&\nabla_\bt \expect{\mathbf{u}_t \sim \pi_\bt}{R(\mb{x}_t,\mathbf{u}_t) \,
			\big| \, \mb{x}_t}\\
		\notag
		&\hskip 0.15in = \nabla_\bt \expect{\pmb{\epsilon}\sim
			\mcal{N}(\mb{0},\mb{I}^\id_d)}{R(\mb{x}_t, \pmb{\Sigma}_\bt \pmb{\epsilon} +
			\pmb{\mu}_\bt) \, \big| \, \mb{x}_t}\\
		\notag
		&\hskip 0.15in = \expect{\pmb{\epsilon}\sim
			\mcal{N}(\mb{0},\mb{I}^\id_d)}{\nabla_\bt R(\mb{x}_t, \pmb{\Sigma}_\bt \pmb{\epsilon} +
			\pmb{\mu}_\bt) \, \big| \, \mb{x}_t }\\
		\label{eq:RewardGrad}
		&\hskip 0.15in = \expect{\pmb{\epsilon}\sim \mcal{N}(\mb{0},\mb{I}^\id_d)}{
			\frac{\partial R_t}{\partial \mathbf{u}_t} \frac{\partial \mathbf{u}_t}{\partial
				\pmb{\Sigma}_\bt} \frac{\partial \pmb{\Sigma}_\bt}{\partial \bt} + \frac{\partial
				R_t}{\partial \mathbf{u}_t} \frac{\partial \mathbf{u}_t}{\partial \pmb{\mu}_\bt}
			\frac{\partial \pmb{\mu}_\bt}{\partial \bt}\, \bigg| \, \mb{x}_t}.
	\end{align}
	Computing $\frac{\partial \mathbf{u}_t}{\partial \pmb{\Sigma}_\bt} \frac{\partial
		\pmb{\Sigma}_\bt}{\partial \bt}$ involves a multiplication of two tensors, which can
	be bypassed by vectorizing $\pmb{\Sigma}_\bt$. It follows that
	\begin{equation*}
		\frac{\partial \mathbf{u}_t}{\partial \pmb{\Sigma}_\bt} \frac{\partial
			\pmb{\Sigma}_\bt}{\partial \bt}  = \frac{\partial \mathbf{u}_t}{\partial \mb{v}_\bt}
		\frac{\partial \mb{v}_\bt}{\partial \bt} = \left( \pmb{\epsilon}^\top \otimes
		\mb{I}^\id_{d} \right) \frac{\partial \mb{v}_\bt}{\partial \bt}.
	\end{equation*}
	On the other hand, $\frac{\partial \mathbf{u}_t}{\partial \pmb{\mu}_\bt}
	\frac{\partial \pmb{\mu}_\bt}{\partial \bt} 
	= \frac{\partial (\pmb{\Sigma}_\bt \pmb{\epsilon}+\pmb{\mu}_\bt)}{\partial
		\pmb{\mu}_\bt} \times
  \frac{\partial \pmb{\mu}_\bt}{\partial \bt} 
	= \frac{\partial \pmb{\mu}_\bt}{\partial \bt} $. This verifies the closed form
	of $\mb{a}_t$ given in the theorem. The closed form of $\mb{B}_t$ can be
	similarly derived by computing the Jacobian of $\mj^\mb{C}(\pi_\bt,\mb{x}_t)$ at
	$\bt=\bt_t$ (with $\mb{c}_t$ simply being the zeroth-order term in the Taylor
	expansion), carrying out the reparametrization trick and then deriving the
	closed form of the partial derivatives.
	
	The first-order term in the Taylor approximation of
	constraint~\eqref{eq:KLDivCondition} vanishes, and the second order term is used
	in the relaxed constraint~\eqref{eq:KLDivRelax}. Matrix $\mb{F}_t$ is the
	\emph{Fisher information matrix} (FIM) that is positive semidefinite by
	construction. 
	As provided in the theorem statement, the closed form of FIM for a Gaussian
	vector is well-known; see e.g., \cite{OB-YIA:2013}. 
	
	\subsection{Proof of Lemma~\ref{lemma:partialDerivative}}
	\label{proofLemma3}
	%  The method of calculation of the partial derivatives follows from the
	%definition of a partial derivative.
	The numerical result proposed in the Lemma arises from the following
	observation.
	\begin{remark}[Evaluating partials from total derivatives]
		For a generic problem, let the independent variable $\mb{p}$ and dependent variable $\mb{q}$ be related
		through the implicit equation $\mb{F}(\mb{p},\mb{q}) = \mb{0}$, where
		$\mb{F}$ is a smooth function. The total
		differential of $\mb{F}(\mb{p},\mb{q}) = \mb{0}$ is given as
		\begin{equation}
			\label{eq:TotalDifferential}
			\left[ \frac{\partial \mb{F}(\mb{p},\mb{q})}{\partial \mb{p}} \right]^\top
			d\mb{p} + \left[ \frac{\partial \mb{F}(\mb{p},\mb{q})}{\partial \mb{q}}
			\right]^\top d\mb{q} = \mb{0}.
		\end{equation}
		To calculate $\frac{\partial p_i}{\partial q_j}$, we set $dp_g = 0$ for all
		$g\neq j$ (a partial derivative with respect to $p_j$ means that any variables
		$p_g$ with $g\neq j$ are assumed to be constant), and
		solve~\eqref{eq:TotalDifferential} for $dp_j$ and $d\mb{q}$. From here, we have
		$\frac{\partial q_i}{\partial p_j} = \frac{d q_i}{d p_j}$.
	\end{remark}
	
		Now, we show that the system of equations~\eqref{eq:totalDeriv} always admits a
		solution when all action variables in $\mb{u}_t$ except one are nullified. We consider the real and imaginary parts of any complex differential equivalent to two real independent differentials. First, we calculate the number of linearly independent equations in~\eqref{eq:totalDeriv}. In order to detect linear dependencies, we note that equations among~\eqref{eq:TD5a}-\eqref{eq:TD9} which only contain a single term of the form $\Re(ds_{i,j})$ or $\Im(ds_{i,j})$ may be combined into~\eqref{eq:TD1}. Using this observation, we note that~\eqref{eq:TD5}, and~\eqref{eq:TD7}-\eqref{eq:TD9} are linearly dependent on~\eqref{eq:TD1}. The total number of linearly independent equations, denoted by $N^\text{eqn}$, is therefore given as
		\begin{align*}
			N^\text{eqn} = H \left( 2|\mcal{N}| + 2|\mcal{E}| + 2|\mcal{N}^\text{ESS}| + 2|\mcal{N}^\text{droop}| \right).
		\end{align*}
		On the other hand, the number of variables in $\mb{u}_t$ is given as
		\begin{align*}
			N^\text{action} = H\left( 2|\mcal{N}^\text{MT}|+2|\mcal{N}^\text{RES}|+3|\mcal{N}^\text{ESS}| + |\mcal{N}^\text{L}|\right),
		\end{align*}
		while the total number of variables in~\eqref{eq:totalDeriv} is given as
		\begin{align*}
			N^\text{vars} = H\big(& 3|\mcal{N}| + 3|\mcal{E}| + 3|\mcal{N}^\text{ESS}| +|\mcal{N}^\text{MT}|\\&  + |\mcal{N}^\text{RES}| + |\mcal{N}^\text{L}| + 1\big).
		\end{align*}
		From the above, it can be verified that $\big( N^\text{vars}-N^\text{action} + 1\big) - N^\text{eqns} > 0$ when $|\mcal{N}^\text{L}| \geq |\mcal{N}^\text{MT}|+|\mcal{N}^\text{RES}|+|\mcal{N}^\text{ESS}|$. Thus, the system of equations~\eqref{eq:totalDeriv}, when all but one action variable differentials are nullified, is a strictly underdetermined system of homogeneous equations. Thus, it has a non-empty nullspace, and therefore a solution always exists.
%	\subsection{Calculation of $\frac{\partial R_t}{\partial \mb{u}_t}$ for multiphase systems}
%		\label{app:mphase}
%		Once the reader is familiar with the total derivative equation~\eqref{eq:totalDeriv} and how it is leveraged by Lemma~\ref{lemma:partialDerivative} to compute $\frac{\partial R_t}{\partial \mb{u}_t}$ in the single phase case, the obvious challenge in multiphase systems emerges with the rank constraint~\eqref{eq:ACDF3}. Unlike the constraint~\eqref{eq:PFcons3} for single phase system, it is difficult to calculate the total derivative of the rank-1 constraint for its multiphase counterpart~\eqref{eq:ACDF3}. The difficulty can be alleviated by introducing an auxiliary variable $\pmb{\nu}_{ij,t}\in\mathbb{C}^6$ and restating the rank-1 constraint as
%		\begin{align*}
%			\bm{\mb{v}_{i,t} & \mb{S}_{ij,t}\\\mb{S}_{ij,t}^H&\mb{l}_{ij,t}} = \pmb{\nu}_{ij,t}\pmb{\nu}_{ij,t}^H.
%		\end{align*}
%		The total derivative of the above equation can be calculated in terms of $d\mb{v}_{i,t}$, $d\mb{S}_{ij,t}$, $d\mb{l}_{ij,t}$, and $d\pmb{\nu}_{ij,t}$. This can be used in combination with Lemma~\ref{lemma:partialDerivative} to calculate the required partial derivatives.
		

	% biography
	\begin{IEEEbiography}[{\includegraphics[width=1in,height=1.25in,clip,keepaspectratio]{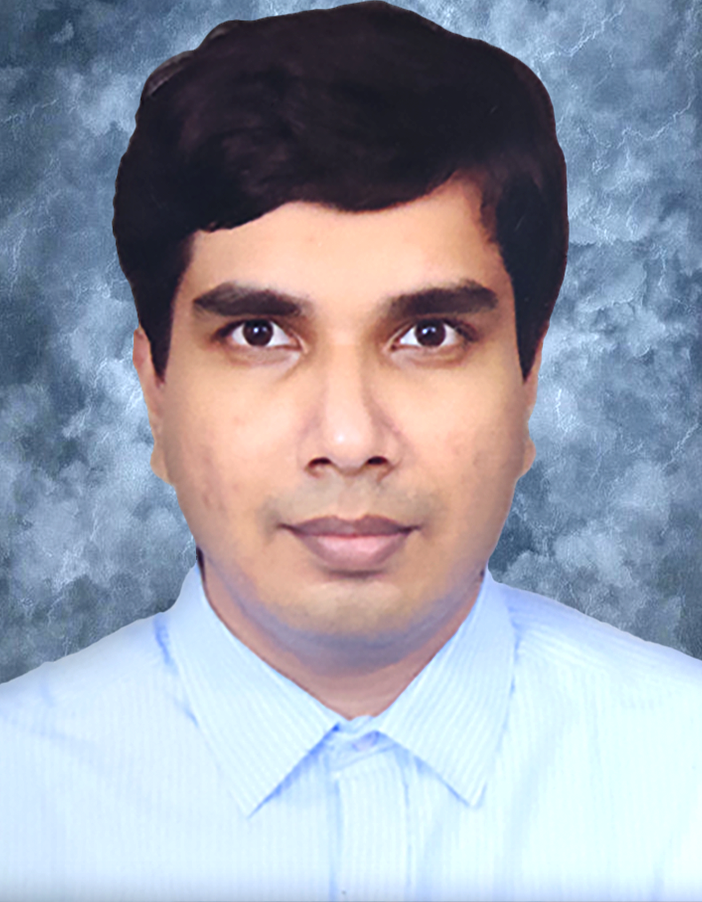}}]%
		{Shourya Bose} (Graduate Student Member, IEEE) received the B.E. and M.Sc. degrees in Electrical and Electronics Engineering and Mathematics from BITS Pilani, KK Birla Goa Campus, India. He is currently working towards Ph.D. degree in Electrical and Computer Engineering at the University of California, Santa Cruz.
		
		He was a co-recipient of the Early Career Best Paper Award given by the Energy, Natural Resources, and the Environment (ENRE) section
		of the Institute of Operations Research and the Management Sciences
		(INFORMS) in 2021.
		
		His research interests involve addressing problems in power systems engineering using tools from optimization theory, machine learning, and control theory.
	\end{IEEEbiography}
	\begin{IEEEbiography}[{\includegraphics[width=1in,height=1.25in,clip,keepaspectratio]{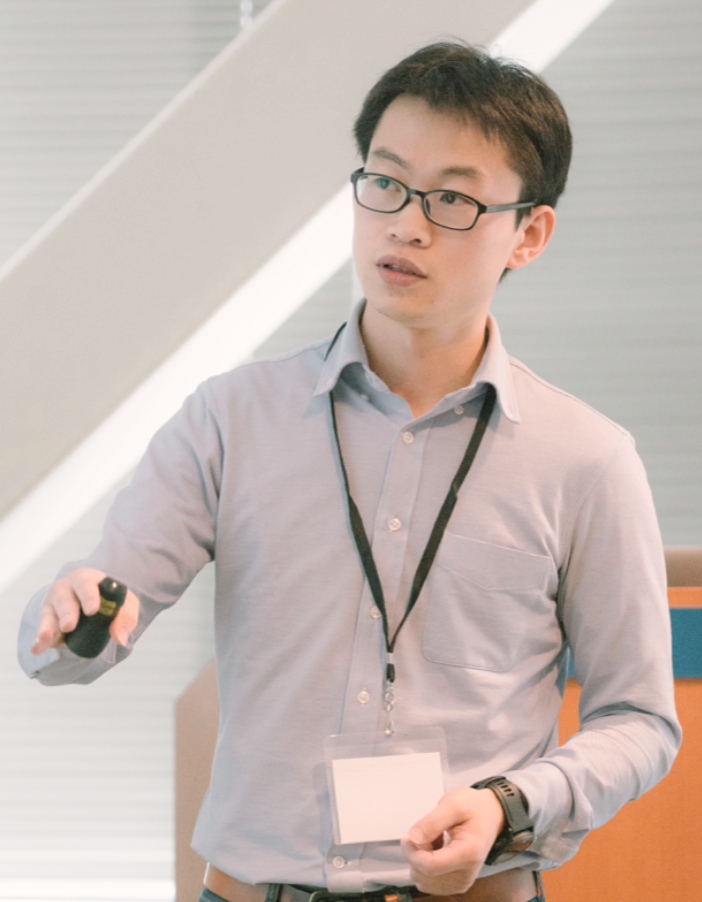}}]%
		{Yu Zhang} (Member, IEEE) received the Ph.D.
		degree in Electrical and Computer Engineering
		from the University of Minnesota, Minneapolis,
		MN, USA, in 2015.
		
		He is currently an Assistant Professor with
		ECE Department, University of California, Santa
		Cruz (UCSC), Santa Cruz, CA, USA. Prior to
		joining UCSC, he was a Postdoc with the University of California, Berkeley, Berkeley, USA,
		and Lawrence Berkeley National Laboratory,
		Berkeley. His research interests include cyberphysical systems, smart power grids, optimization theory, machine
		learning, and big data analytics.
		
		Dr. Zhang was the recipient of the Hellman Fellowship in 2019. He
		was the co-recipient of the Early Career Best Paper Award given by
		the Energy, Natural Resources, and the Environment (ENRE) section
		of the Institute of Operations Research and the Management Sciences
		(INFORMS) in 2021.
	\end{IEEEbiography}
	
\end{document}